\documentclass{article}%
\usepackage{amsmath, amsthm, amssymb}
\usepackage{pslatex}
\usepackage{amsfonts}
\usepackage{verbatim}
\usepackage{epsfig, graphicx}
\usepackage{amsmath}
\usepackage{amssymb}
\usepackage{graphicx}%
\theoremstyle{plain}
\newtheorem{theorem}{Theorem}

\newtheorem{defn}{Definition}

\newtheorem{lem}{Lemma}

\newtheorem{prop}{Proposition}
\theoremstyle{remark}

\DeclareMathOperator{\Diag}{diag}

\begin{document}

\title{Approximating multi-dimensional Hamiltonian flows by billiards.}
\author{A. Rapoport\thanks{Weizmann Institute of Science. E-mail: \textsl{anna.rapoport@weizmnan.ac.il}.},~~V. Rom-Kedar\thanks{Weizmann Institute of
Science. E-mail:
\textsl{vered.rom-kedar@weizmnan.ac.il}.}~~and~~D. Turaev
\thanks{Ben-Gurion University. E-mail: \textsl{turaev@cs.bgu.ac.il}.}}


\date{\today}
\maketitle

\begin{abstract}
Consider a family of smooth potentials $V_{\epsilon}$, which, in the limit
$\epsilon\rightarrow0$, become a singular hard-wall potential of a
multi-dimensional billiard. We define auxiliary billiard domains that
asymptote, as $\epsilon\rightarrow0$ to the original billiard, and provide
asymptotic expansion of the smooth Hamiltonian solution in terms of these
billiard approximations. The asymptotic expansion includes error estimates in
the $C^{r}$ norm and an iteration scheme for improving this approximation.
Applying this theory to smooth potentials which limit to the multi-dimensional
close to ellipsoidal billiards, we predict when the separatrix splitting
persists for various types of potentials.

\end{abstract}

\section{Introduction}

Imagine a point particle travelling freely (without friction) on a table,
undergoing elastic collisions with the edges of the table. The table is just a
bounded region of the plane. This model resembles a game of billiards, but it
looks much simpler - we have only one ball, which is a dimensionless point
particle. There is no friction and the table has no pockets. The shape of the
table determines the nature of the motion (see \cite{KzTr91} and references
therein) - it can be ordered (integrable, e.g. in ellipsoidal tables), ergodic
(e.g. in generic polygons), strongly mixing (in dispersing-Sinai tables or
focusing-Bunimovich tables), or of a mixed nature for a general geometry with
both concave and convex boundary components. A mechanical realization of this
model in higher dimensions appears when one considers the motion of $N$ rigid
$d$-dimensional balls in a $d$-dimensional box ($d=2$ or $3$): it corresponds
to a billiard problem in a complicated $n$ dimensional domain, where
$n=2N\times d$ (\cite{Sina63,ChMar03}).

Usually, in the physics context, this billiard description is used to model a
more complicated flow by which a particle is moving approximately inertially,
and then is reflected by a steep potential. The reduction to the billiard
problem simplifies the analysis tremendously, often allowing to describe
completely the dynamics in a given geometry. Numerous applications of this
idea appear in the physics literature; It works as idealized model for the
motion of charged particles in a steep potential, a model which is often used
to examine the relation between classical and quantized systems (see
\cite{Gut90,Smil95} and references therein); This approximation was utilized
to describe the dynamics of the motion of cold atoms in dark optical traps
(see \cite{kfad01} and reference therein); This model has been suggested as a
first step for substantiating the basic assumption of statistical mechanics --
the ergodic hypothesis of Boltzmann (\cite{Kry79},\cite{Sina63},\cite{Sina70}%
,\cite{SiCh87},\cite{Sz96}). The opposite point of view may be taken when one
is interested in studying numerically the hard wall system in a complicated
geometry (e.g. apply ideas of \cite{Mar68} to \cite{MarW01}) - then designing
the "correct" limiting smooth Hamiltonian may simplify the complexity of the programming.

For two-dimensional finite-range axis-symmetric potentials
\cite{Sina63,Ku76,KuMu81,Bal88,Kn89,DoLi91,Do96,BaTo04}, it was shown that a
modified billiard map may be defined, and several works have utilized this
modified map to prove ergodicity of some configurations
\cite{Sina63,Ku76,KuMu81,DoLi91,BaTo04}, or to prove that other configurations
may possess stability islands \cite{Bal88,Do96}. The general problem of
studying the limiting process of making a steep two-dimensional potential
steeper up to the hard-wall limit can be approached in a variety of ways. In
\cite{Mar68} approach based on generalized functions was proposed. In
\cite{TuRK98} we developed a different paradigm for studying this problem. We
first formulated a set of conditions on general smooth steep potentials in
two-dimensional domains ($C^{r}$ smooth, not necessarily of finite range, nor
axis-symmetric) which are sufficient for proving that regular reflections of
the billiard flow and of the smooth flow are close in \textit{the $C^{r}$
topology}. This statement, which may appear first as a mathematical exercise,
is quite powerful. It allows to prove immediately the persistence of various
kinds of billiard orbits in the smooth flows (see \cite{TuRK98} and Theorem
\ref{thm:persistance} in Section \ref{sec:persis}) and to investigate the
behavior near singular orbits (e.g. orbits which are tangent to the boundary)
by combining several Poincare maps, see for example
\cite{RKTu99,turk03,Chen04}. The first part of this paper (see Theorems
\ref{Main th0}-\ref{Main thr}) is a generalization of this result to the
multi-dimensional case.

Thus, it appears that the Physicists approach, of approximating the smooth
flow by a billiard has some mathematical justification. How good is this
approximation? Can this approach be used to obtain an asymptotic expansion to
the smooth solutions? The second part of this paper answers these questions.
We propose an approximation scheme, with a constructive twist - we show that
the best zero-order approximation should be a billiard map in a slightly
distorted domain. We provide the scaling of the width of the corresponding
boundary layer with the steepness parameter and with the number of derivatives
one insists on approximating. Furthermore, the next order correction is
explicitly found, supplying a modified billiard map (reminiscent of the
shifted billiard map of \cite{Sina63,Do96}) which may be further studied. We
believe this part is the most significant part of the paper as it supplies a
constructive tool to study the difference between the smooth flow and the
billiard flow.

Indeed, in the last part of this paper we demonstrate how these tools may be
used to instantly extend novel results which were obtained for billiards to
the steep potential setting; It is well known that the billiard map is
integrable inside an ellipsoid \cite{KzTr91}. Moreover, Birkhoff-Poritski
conjecture claims that in 2 dimensions among all the convex smooth concave
billiard tables only ellipses are integrable \cite{Tab95}. In \cite{Ves91}
this conjecture was generalized to higher dimensions. Delshams\textit{\ et al}
(\cite{DFRR01}, \cite{DeRa96} see references therein) studied the affect of
small entire symmetric perturbations to the ellipsoid shape on the
integrability. They proved that in some cases the separatrices of a simple
periodic orbit split; Thus, they proved a local version of Birkhoff conjecture
in the 2 dimensional setting, and provided several non-integrable models in
the $n$ dimensional case. Here, we show that a simple combination of their
results with ours, extends their result to the smooth case - namely it shows
that the Hamiltonian flow, in a sufficiently steep potential which
asymptotically vanishes in a shape which is a small perturbation of an
ellipsoid, is chaotic. Furthermore, we quantify, for a given perturbation of
the ellipsoidal shape, what \textquotedblleft sufficiently steep" means for
exponential, Gaussian and power-law potentials.

These results may give the impression that the smooth flow and the billiard
flow are indeed very similar, and so a Scientist's dream of greatly
simplifying a complicated system is realized here. In the discussion we go
back to this point - as usual dreams never materialize in full.

The paper is ordered as follows; In Section \ref{sec:bil} we define and
describe the billiard flow and billiard map. In Section \ref{sec:smooth} we
study the smooth Hamiltonian flow; we first prove that if the potential
satisfies some natural conditions the smooth regular reflections will limit
smoothly to the billiard's regular reflections (Theorems \ref{Main th0}%
,\ref{Main thr}). Then, we define a natural Poincar\'{e} section on which a
generalized billiard map may be defined for the smooth flow. Next, we derive
the correction term to the zeroth order billiard approximation (Theorem
\ref{thm-est}) and calculate it for three model potentials (exponential,
Gaussian and power-law). We end this section by stating its immediate
implication - a persistence theorem for various types of trajectories (Theorem
\ref{thm:persistance}). In Section \ref{sec:ellips} we apply these results to
the perturbed ellipsoidal billiard. We end the paper with a short summary and
discussion. The appendices contain most of the proofs, whereas in the body of
the paper we usually only indicate their main steps.

\section{Billiards in $d$ dimensions}

\label{sec:bil}

\subsection{The Billiard Flow}

Consider a billiard flow as the motion of a point mass in a compact domain
$D\in\mathbb{R}^{d}$ or $\mathbb{T}^{d}$. Assume that the boundary $\partial
D$ consists of a finite number of $C^{r+1}$ smooth ($r\geq1$) $(d-1)$%
-dimensional submanifolds:%
\begin{equation}
\partial D=\Gamma_{1}\cup\Gamma_{2}\cup...\cup\Gamma_{n},\;\;\;\;\;i=1\ldots
n. \label{Domain boundary}%
\end{equation}
The boundaries of these submanifolds, when exist, form \emph{the corner set}
of $\partial D$:%
\begin{equation}
\Gamma^{\ast}=\partial\Gamma_{1}\cup\partial\Gamma_{2}\cup...\cup
\partial\Gamma_{n},\;\;\;\;\;i=1\ldots n. \label{Corner set}%
\end{equation}
The moving particle has a position $q\in D$ and a momentum vector
$p\in\mathbb{R}^{d}$ which are functions of time. If $q\in int(D)$, then the
particle moves freely with the constant velocity according to the
rule\footnote{We assume that the particle has mass one (otherwise rescale
time).}:
\begin{equation}
\left\{
\begin{array}
[c]{l}%
\dot{q}=p\\
\overset{\cdot}{p}=0
\end{array}
\right.  . \label{free motion}%
\end{equation}
Equation (\ref{free motion}) is Hamiltonian with the Hamiltonian function
(hereafter $p^{2}=\langle p,p\rangle$)%
\begin{equation}
H(q,p)=\frac{p^{2}}{2}. \label{free Hamiltonian}%
\end{equation}
The particle moves at a constant speed and bounces of $\partial D$ according
to the usual elastic reflection law : \emph{the angle of incidence is equal to
the angle of reflection}. This means that the outgoing vector $p_{out}$ is
related to the incoming vector $p_{in}$ by%
\begin{equation}
p_{out}=p_{in}-2\langle p_{in},n(q)\rangle n(q), \label{Reflection}%
\end{equation}
where $n(q)$ is the inward unit normal vector to the boundary $\partial D$ at
the point $q$, see \cite{ChMar03}. To use the reflection rule
(\ref{Reflection}), we need the normal vector $n(q)$ to be defined, hence the
rule cannot be applied at points $q\in\Gamma^{\ast}$, where such a vector
fails to exist\footnote{To be precise, one may define $n(q)$ by continuity at
points of $\Gamma^{\ast}$, but this might give more than one normal vector
$n(q)$, hence the dynamics would be multiply defined for a generic corner. We
adopt a standard convention that the reflection is not defined at any
$q\in\Gamma^{\ast}$.}.

\begin{defn}
\textrm{The domain $D$ is called the \emph{configuration space} of the
billiard system. }
\end{defn}

The phase space of the system is $\mathcal{P}=D\times S^{d-1}$, where
$S^{d-1}$ is a $(d-1)$-dimensional unit sphere (we set $H=\frac{1}{2}$) of
velocity vectors. So the elements of $\mathcal{P}$ are
\[
\rho\equiv(q,p).
\]
Denote the time $t$ map of the billiard flow as
\begin{equation}
b_{t}:\rho_{0}\rightarrow\rho_{t}.
\end{equation}
We do not consider reflections at the points of the corner set, so $\rho
_{t}=b_{t}\rho_{0}$ implies here that the distance between any point on the
trajectory connecting $q_{0}$ with $q_{t}$ and the set $\Gamma^{\ast}$ is
bounded away from zero. A point $\rho\in\mathcal{P}$ is called \emph{an inner
point} if $q\notin\partial D$ and \emph{a collision point} if $q\in\partial
D\setminus\Gamma^{\ast}$. Obviously, if $\rho_{0}$ and $\rho_{t}=b_{t}\rho
_{0}$ are inner points, then $\rho_{t}$ depends continuously on $\rho_{0}$ and
$t$. If $\rho_{t}$ is a (non-tangent) collision point then the velocity vector
undergoes a jump. Thus, in this case both $b_{t-0}$ and $b_{t+0}$ are defined.
The map $R_{\circ}=b_{t+0}b_{t-0}^{-1}$ is the reflection law
(\ref{Reflection}) (augmented by $q_{out}=q_{in}$).

\begin{figure}[ptb]
\centering \psfig{figure =
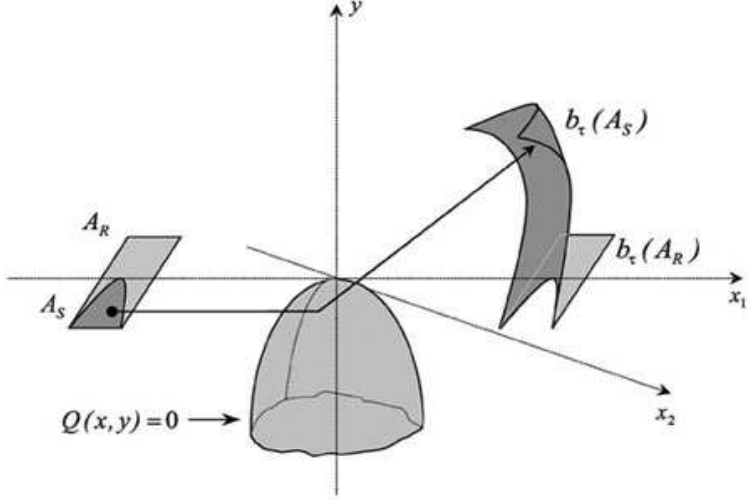,height=70mm,width=100mm}\caption{Singularity near a
tangent trajectory. For better visualization we present a slanted hyperplane
which is divided into 2 parts: $b_{\tau}$ has a square-root singularity on the
boundary between $A_{R}$ and $A_{S}$.}%
\label{fig:singular_tangency}%
\end{figure}

If the piece of trajectory that connects $q_{0}$ with $q_{t}$ does not have
tangencies with the boundary, then $\rho_{t}$ depends $C^{r}$-smoothly on
$\rho_{0}$. It is well-known (\cite{Sina70},\cite{TuRK98}) that the map
$b_{t}$ loses smoothness at any point $q_{0}$ whose trajectory is tangent to
the boundary at least once on the interval $(0,t)$. Clearly a tangency may
occur only if the boundary is concave in the direction of motion at the point
of tangency. Consider hereafter only \emph{non-degenerate tangencies}, namely
assume that the curvature in the direction of motion does not vanish.

Choose local coordinates $q=(x,y)$ in such a way that the origin corresponds
to the collision point, the $y$-axis is normal to the boundary and looking
inside the billiard region $D$, and the $x$-coordinates ($x\in\mathbb{R}%
^{d-1}$) correspond to the directions tangent to the boundary. If $Q(x,y)=0$
is the equation of the boundary in these coordinates, then $Q_{y}(0,0)\neq0$
and $Q_{x}(0,0)=0$. We choose the convention that $Q_{y}(0,0)>0$. Obviously,
the tangent trajectory is characterized by the condition $p_{y}=0$, where
$(p_{x},p_{y})$ are the components of the momentum $p$. The vector $p_{x}%
=\dot{x}$ indicates the direction of motion of the tangent trajectory. It is
easy to check that the tangency is non-degenerate if and only if%
\begin{equation}
p_{x}^{T}Q_{xx}(0,0)p_{x}>0. \label{nondegtan}%
\end{equation}
Notice that if the billiard's boundary has saddle points (or if the billiard
is semi-dispersing), then there always exist directions for which this
non-degeneracy assumption fails. On the other hand, if the boundary is
strictly concave, then all tangencies are non-degenerate.

Let $x=(x_{1},\dots,x_{d-1})$ with $x_{1}$ corresponding to the direction of
motion (i.e. $p_{x}=(1,0,\dots,0)$). Then, the boundary surface near the point
of non-degenerate tangency is described by the following equation:%
\[
y=-\alpha x_{1}^{2}+O(z^{2},x_{1}z),\qquad\alpha>0,
\]
where we denote $z=(x_{2},\dots,x_{d-1})$. It is easy to see now that for a
non-degenerate tangency, for a small $\tau$ the map $b_{\tau}$ of the line
$\rho_{0}=(x_{0}=(-\tau/2,0,\dots,0),\;y_{0}\leq0,\;p_{0x}=(1,0,\dots
,0),\;p_{0y}=0)$ is given by%
\[
\rho_{\tau}=((\tau/2,0,\dots,0)+O(y_{0}),\ 2\tau\sqrt{-\alpha y_{0}}%
+O(y_{0}),\ (1,0,\dots,0)+O(y_{0}),\ 4\sqrt{-\alpha y_{0}}+O(y_{0})).
\]
As we see, the billiard flow looses smoothness indeed (it has a square-root
singularity in the limit $y\rightarrow-0$) near the tangent trajectory. See
Figure \ref{fig:singular_tangency}.

\subsection{The Billiard map\label{bmd}}

It is standard in dynamical system theory to reduce the study of flows to maps
by constructing a cross-section. The latter is a hypersurface transverse to
the flow. For the flow $b_{t}$, such a hypersurface in phase space
$\mathcal{P}$ can be naturally constructed with the help of the boundary of
$D$, i.e. the natural cross-section $S$ corresponds exactly to the collision
points of the flow with the domain's boundary:
\begin{equation}
S=\{\rho=(q,p)\in\mathcal{P}:q\in\partial D,\langle p,n(q)\rangle\geq0\}.
\label{cross-section}%
\end{equation}
This is a $(2d-2)$-dimensional submanifold in $\mathcal{P}$. Any trajectory of
the flow $b_{t}$ crosses $S$ every time it reflects at $\partial D$. This
defines the \emph{Poincar\'{e} map}%
\begin{equation}
B:S\rightarrow S\;\;\mathrm{such\;that}\;\;B\rho=b_{\tau_{\circ}(\rho)+0}\rho,
\label{billiard map}%
\end{equation}
where
\[
\tau_{\circ}(\rho)=\min\{t>0:b_{t+0}\rho\in S\}.
\]

\begin{defn}
\textrm{The map $B$ is called \emph{the billiard map}. }
\end{defn}

It is convenient to represent the billiard map as a composition of a
free-flight and a reflection:%
\[
B=R_{\circ}\circ F_{\circ},
\]
where the free-flight map is given by%
\begin{equation}
F_{\circ}(q,p)=b_{\tau_{\circ}(\rho)-0}(q,p), \label{frbil}%
\end{equation}
and the reflection law is given by%
\[
R_{\circ}(q,p)=(q,p-2\langle p,n(q)\rangle n(q)).
\]
The billiard map $B$ is a $C^{r}-$diffeomorphism at all points $\rho\in
S\setminus\Sigma$ such that $B\rho\in S\setminus\Sigma$, where $\Sigma$ is
\emph{the singular set}
\begin{equation}
\Sigma=\Sigma_{tangencies}\bigcup\Sigma_{corners}=\{(q,p)\in P:\langle
p,n(q)\rangle=0\}\cup\{(q,p)\in P:q\in\Gamma^{\ast}\}, \label{singular set}%
\end{equation}
and $B$ is $C^{0}$ at the non-degenerate tangent trajectories.

\section{\label{sec:smooth}Smooth Hamiltonian approximation}

\subsection{Setup and Conditions on Potential}

Consider the family of Hamiltonian systems associated with:%
\begin{equation}
H=\frac{p^{2}}{2}+V(q;\epsilon), \label{Hamiltonian epsilon}%
\end{equation}
where the $C^{r+1}$-smooth potential $V(q;\epsilon)$ tends to zero inside a
region $D$ as $\epsilon\rightarrow0$, and it tends to infinity (or to a
constant larger than the fixed considered energy level, say $H=\frac{1}{2}$)
outside. Formally, the billiard flow in $D$ may be expressed as a limiting
Hamiltonian system of the form:%
\begin{equation}
H_{b}=\frac{p^{2}}{2}+V_{b}(q), \label{Hamiltonian billiard}%
\end{equation}
where%
\begin{equation}
V_{b}(q)=\left\{
\begin{array}
[c]{ll}%
0\;\;\;\;\;\;\; & q\in D\\
+\infty\;\;\; & q\notin D
\end{array}
\right.  . \label{Potential billiard}%
\end{equation}
Let us formulate conditions under which this simplified billiard motion
approximates the smooth Hamiltonian flow. In the two-dimensional case these
conditions were introduced in \cite{TuRK98}.

\textbf{Condition I.}\emph{\ For any compact region $K\subset D$ the potential
$V(q;\epsilon)$ diminishes along with all its derivatives as $\epsilon
\rightarrow0$:}
\begin{equation}
\lim_{\epsilon\rightarrow0}\Vert V(q;\epsilon)|_{q\in K}\Vert_{C^{r+1}}=0.
\label{Potential diminishes}%
\end{equation}
\indent The growth of the potential to infinity across the boundary needs to
be treated more carefully. We assume that $V$ is evaluated along the level
sets of some \emph{finite} function near the boundary. In other words,
suppose, that in a neighborhood $\tilde{D}$ of $D\backslash\Gamma^{\ast}$
there exists \emph{a pattern function} $Q(q;\epsilon):\tilde{D}\rightarrow
\mathbb{R}^{1}$ which is $C^{r+1}$ with respect to $q$ and it depends
continuously on $\epsilon$ (in the $C^{r+1}$-topology) at $\epsilon\geq0$ (so
it has, along with all derivatives, a proper limit as $\epsilon\rightarrow0$).
See Figure \ref{fig:pattern_function}. Assume that away from $\Gamma^{\ast}$:

\begin{figure}[ptb]
\centering \psfig{figure =
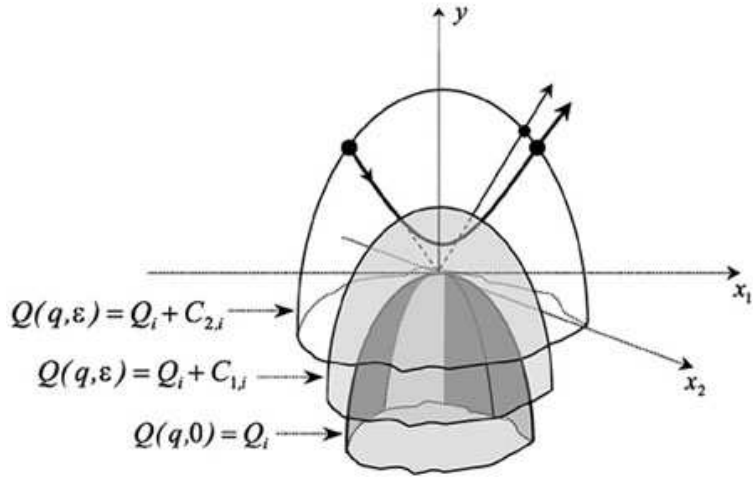,height=70mm,width=100mm}\caption{Level sets of a pattern
function $Q(q;\epsilon)$. A bold line is a trajectory of the Hamiltonian flow
near the boundary; a solid is a billiard trajectory.}%
\label{fig:pattern_function}%
\end{figure}

\begin{figure}[ptb]
\centering \psfig{figure =
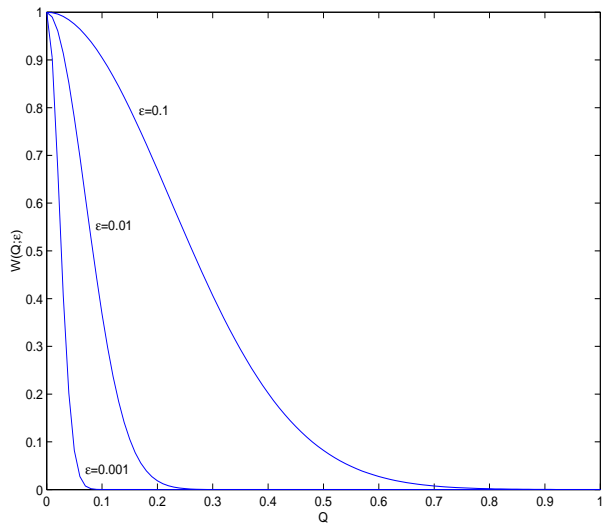,height=70mm,width=80mm}\caption{Gaussian potential given
near the boundary by $W(Q;\epsilon)=e^{-\frac{Q^{2}}{\epsilon}}$ satisfies
Conditions I-IV.}%
\label{fig:pattern_barrier}%
\end{figure}

\textbf{Condition IIa. }\emph{The billiard boundary is composed of level
surfaces of $Q(q;0)$: }%
\begin{equation}
Q(q;\epsilon=0)|_{q\in\Gamma_{i}}\equiv Q_{i}=\mathrm{constant}.
\label{Pattern function}%
\end{equation}

For each neighborhood of the boundary component $\Gamma_{i}$ (so
$Q(q;\epsilon)$ is close to $Q_{i}$), let us define \emph{a barrier function}
$W_{i}(Q;\epsilon):\mathbb{R}^{1}\rightarrow\mathbb{R}^{1}$, which does not
depend explicitly on $q$, and assume that:

\textbf{Condition IIb.} \emph{There exists a small neighborhood $N_{i}$ of the
surface $\Gamma_{i}$ in which:}%
\begin{equation}
V(q;\epsilon)|_{q\in N_{i}}\equiv W_{i}(Q(q;\epsilon)-Q_{i};\epsilon),
\label{Barrier function}%
\end{equation}
and

\textbf{Condition IIc.}$\nabla V$\emph{\ does not vanish in a finite
neighborhood of the boundary surfaces, thus: }%
\begin{equation}
\nabla Q|_{q\in N_{i}}\neq0 \label{Gradient Q}%
\end{equation}
\emph{and}%
\begin{equation}
\frac{d}{dQ}W_{i}(Q-Q_{i};\epsilon)\neq0. \label{Derivative of W}%
\end{equation}

Now, the rapid growth of the potential across the boundary may be described in
terms of the barrier functions alone. Note that by (\ref{Gradient Q}), the
pattern function $Q$ is monotonic across $\Gamma_{i}$, so either $Q>Q_{i}$
corresponds to the points near $\Gamma_{i}$ inside $D$ and $Q<Q_{i}$
corresponds to the outside, or vice versa.

\textbf{Condition III.} \emph{There exists a constant (may be infinite)}%
$\mathcal{E}>0$ \emph{such that as $\epsilon\rightarrow+0$ the barrier
function increases from zero to }$\mathcal{E}$ \emph{ across the boundary
}$\Gamma_{i}$\emph{:}%
\begin{equation}
\lim_{\epsilon\rightarrow+0}W(Q;\epsilon)=\left\{
\begin{array}
[c]{ll}%
0, & Q\text{~~inside~~}D\\
\mathcal{E}, & Q\text{~~outside~~}D
\end{array}
\right.  . \label{Growth of barrier function}%
\end{equation}

By (\ref{Derivative of W}) $Q$ could be considered as a function of $W$ and
$\epsilon$ near the boundary: $Q=Q_{i}+\mathcal{Q}_{~~i}(W;\epsilon)$.
Condition IV states that for small $\epsilon$ a finite change in $W$
corresponds to a small change in $Q$:

\textbf{Condition IV.} \emph{As $\epsilon\rightarrow+0$, for any fixed $W_{1}$
and $W_{2}$ such that $0<W_{1}<W_{2}<\mathcal{E}$, for each boundary component
$\Gamma_{i}$, the function }$\mathcal{Q}_{~~i}(W;\epsilon)$ \emph{tends to
zero uniformly on the interval} $[W_{1},W_{2}]$ \emph{along with all its
}$(r+1)$ \emph{derivatives.}

Figure \ref{fig:pattern_barrier} shows the geometric interpretation of the
pattern function and a typical dependence of the barrier function on $Q$ and
$\epsilon$.

\emph{Note that the use of the pattern and barrier functions essentially
reduces the }$d$\emph{-dimensional Hamiltonian dynamics to a }$1$%
\emph{-dimensional one, which allows for a direct asymptotic integration of
the smooth problem. }

\subsection{C$^{0}$ and C$^{r}$ - closeness Theorems}

\begin{theorem}
\label{Main th0} Let the potential $V(q;\epsilon)$ in
(\ref{Hamiltonian epsilon}) satisfy Conditions I-IV stated above. Let
$h_{t}^{\epsilon}$ be the Hamiltonian flow defined by
(\ref{Hamiltonian epsilon}) on an energy surface $H=H^{\ast}<\mathcal{E}$, and
$b_{t}$ be the billiard flow in $D$. Let $\rho_{0}$ and $\rho_{T}=b_{T}%
\rho_{0}$ be two inner phase points\footnote{Hereafter, $T$ always denotes
a\emph{ finite }number.}. Assume that on the time interval $[0,T]$ the
billiard trajectory of $\rho_{0}$ has a finite number of collisions, and all
of them are either regular reflections or non-degenerate tangencies. Then
$h_{t}^{\epsilon}\rho_{\overset{\longrightarrow}{_{\epsilon\rightarrow0}}%
}b_{t}\rho$, uniformly for all $\rho$ close to $\rho_{0}$ and all $t$ close to
$T$.
\end{theorem}

\begin{theorem}
\label{Main thr} In the conditions of Theorem \ref{Main th0}, further assume
that the billiard trajectory of $\rho_{0}$ has no tangencies to the boundary
on the time interval $[0,T]$. Then $h_{t}^{\epsilon}~_{\overset
{\longrightarrow}{_{\epsilon\rightarrow0}}} b_{t}$ in the $C^{r}$-topology in
a small neighborhood of $\rho_{0}$, and for all $t$ close to $T$.
\end{theorem}

The proof of the theorems is presented in the appendix and it follows closely
the proof in \cite{TuRK98}. Informally, the logic behind Conditions I-IV is as follows.

Condition I, obviously, implies that the particle moves with almost constant
velocity (along a straight line) in the interior of $D$ until it reaches a
thin layer near the boundary where $V$ runs from zero to large values (a
smaller $\epsilon$ corresponds to a thinner boundary layer). Note that the
boundary layer can not be fully penetrated by the particle. Indeed, as in all
mechanical Hamiltonians, the energy level defines the region of allowed
motion: for a fixed energy level $H=H^{\ast}<\mathcal{E}$, all trajectories
stay in the region $V(q;\epsilon)\leq H^{\ast}$. It follows from Condition III
that for any such $H^{\ast}$, the region of allowed motion approaches $D$ as
$\epsilon\rightarrow0$. Thus, by Condition III, if the particle enters the
layer near a boundary surface (note that points from $\Gamma^{\ast}$ are not
considered in this paper), it has, in principle, two possibilities. First, it
may be reflected and then exits the boundary layer near the point it entered.
The other possibility, which we want to avoid, is that the particle sticks to
the boundary and travels along it far from the entrance point. Condition IV
guarantees that if the reflection is regular, or in case of non-degenerate
tangency, the travel distance along the boundary vanishes asymptotically with
$\epsilon$. The case of degenerate tangencies, which are unavoidable in the
higher dimensional case if the boundary has directional curvatures of opposite
signs (namely saddle points), is not studied here. Once we know that the time
spent by the particle near the boundary is small, we can see that Condition II
guarantees that the reflection will be of the right character, namely the
smooth reflection is $C^{0}$-close to that of the billiard. Indeed, Condition
II implies that the reaction force is normal to the boundary, hence, as the
time of collision is small and the position of the particle does not change
much during this time, the direction of the force stays nearly constant during
the collision. Thus, only the normal component of the momentum is changing
sign while the tangent components are nearly preserved. Computations along
these lines provide a proof of Theorem \ref{Main th0}.

Proving Theorem \ref{Main thr}, i.e. the $C^{r}$-closeness, makes a
substantial use of Condition IV. Let us explain in more detail the difference
between the $C^{0}$ and $C^{r}$ topologies in this context. Take the same
initial condition $(q_{0},p_{0})$ for a billiard orbit and for an orbit of the
Hamiltonian system (\ref{Hamiltonian epsilon}) (the Hamiltonian orbit will be
called \textit{the smooth orbit}). Consider a time interval $t$ for which the
billiard orbit collides with the boundary only once. In these notations
$\varphi_{in}$ is the angle between $p_{0}$ (the momentum at the point $q_{0}%
$) and the normal to the boundary at the collision point, $\varphi_{out}$ is
the angle between $p_{t}$ (the velocity vector at the point $q_{t}$) and the
normal. Define the incidence and reflection angles ($\varphi_{in}(\epsilon)$
and $\varphi_{out}(\epsilon)$) for the smooth trajectory in the same way.
Theorem \ref{Main th0} implies the correct reflection law for smooth
trajectories:%
\begin{equation}
\varphi_{in}(\epsilon)+\varphi_{out}(\epsilon)\approx0
\label{Angels reflection}%
\end{equation}
for sufficiently small $\epsilon$. However, $\varphi_{in}+\varphi_{out}$ is a
function of the initial conditions, so a non-trivial question is when it is
close to zero along with all its derivatives. In Theorem \ref{Main thr} we
prove that Condition IV is sufficient for guaranteeing the correct reflection
law in the $C^{r}$-topology in the case of non-tangent collision (near tangent
trajectories the derivatives of the smooth flow cannot converge to those of
the billiard because the billiard flow is singular there, see Figure
\ref{fig:singular_tangency}).

Hereafter, we will fix the energy level of the Hamiltonian flow to $H^{\ast
}=\frac{1}{2}$. Notice that the analysis may be applied to systems with steep
potentials which do not depend explicitly on $\epsilon$ (or do not degenerate
as $\epsilon\rightarrow0$) in the limit of sufficiently high energy: the
reduction to the setting (\ref{Hamiltonian epsilon}) which we consider here
may be achieved by a scaling of time.

\subsection{\label{sec:crestimates}Asymptotic for a regular reflection}

It follows from the proof of Theorem \ref{Main thr} that the behavior of
smooth trajectories close to billiard trajectories of regular reflections can
be described by an analogue of the billiard map. More precisely, one can
construct a cross-section $S_{\epsilon}$ in phase space of the Hamiltonian
flow, close to the \textquotedblleft natural\textquotedblright\ cross-section
$S$ where the billiard map $B$ is defined; the trajectories of the Hamiltonian
flow which are close to regular billiard trajectories define the Poincar\'{e}
map on $S_{\epsilon}$, and this map is $C^{r}$-close to $B$. Let us explain
this in more details.

\begin{figure}[ptb]
\centering \psfig{figure =
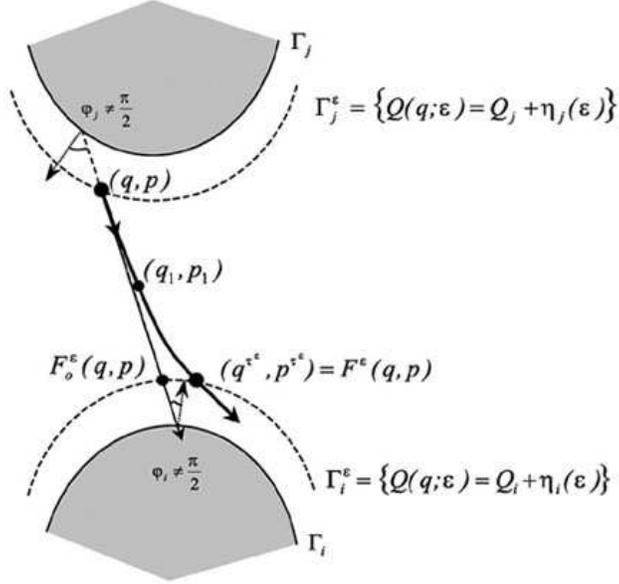,height=80mm,width=85mm}\caption{Free flight between boundaries
$\Gamma_{i}^{\epsilon}$ and $\Gamma_{j}^{\epsilon}$. A smooth trajectory is
marked by a bold line and an auxiliary billiard trajectory is marked by a
solid line.}%
\label{fig:free flight}%
\end{figure}

It is convenient to consider an auxiliary billiard in the modified domain
$D^{\epsilon}$, defined as follows. For each boundary surface $\Gamma_{i}$,
take any $\nu_{i}(\epsilon)\rightarrow+0$ such that the function (inverse
barrier) $\mathcal{Q}_{~~i}(W;\epsilon)$ tends to zero along with all its
derivatives, uniformly for $\frac{1}{2}\geq W\geq\nu_{i}$. We will use the
notation
\begin{equation}
M_{i}^{(r)}(\nu;\epsilon)=\sup_{{\footnotesize
\begin{array}
[c]{c}%
\nu\leq W\leq\frac{1}{2}\\
0\leq l\leq r+1
\end{array}
}}|\mathcal{Q}_{\;i}^{(l)}(W;\epsilon)|. \label{M-big}%
\end{equation}
Condition IV implies that $M$ approaches zero as $\epsilon\rightarrow0$ for
any fixed $\nu>0$, hence the same holds true for any sufficiently slowly
tending to zero $\nu(\epsilon)$, i.e. the required $\nu_{i}(\epsilon)$ exist.
Let $\eta_{i}(\epsilon)=\mathcal{Q}_{~~i}(\nu_{i};\epsilon)$ and consider the
billiard in the domain $D^{\epsilon}$ which is bounded by the surfaces
$\Gamma_{i}^{\epsilon}:{Q(q;\epsilon)=Q_{i}+\eta_{i}(\epsilon)}$. See Figure
\ref{fig:free flight}. Recall that the boundaries $\Gamma_{i}$ of the original
billiard table $D$ are level sets $Q(q;0)=Q_{i}$, and that $\eta_{i}%
(\epsilon)\rightarrow0$ by construction, so the new billiard is close to the
original one. In particular, for regular reflections, the billiard map
$B^{\epsilon}$ of the auxiliary billiard tends to the original billiard map
$B$ along with all its derivatives. It is established in the proof of Theorem
\ref{Main thr} that for any choice of $\nu_{i}$'s tending to zero, the
condition $q\in\partial D^{\epsilon}$ defines a cross-section in the phase
space of the smooth Hamiltonian flow; Trajectories which are close to the
billiard trajectories of regular reflection, i.e. those which intersect
$\partial D^{\epsilon}$ at an angle bounded away from zero, define the map
\[
F^{\epsilon}:(q\in\partial D^{\epsilon}%
,p\;\;\mbox{ looking inwards }\;D^{\epsilon})\rightarrow(q\in\partial
D^{\epsilon},p\;\;\mbox{ looking
outwards }\;D^{\epsilon}),
\]
namely%
\begin{equation}
F^{\epsilon}(q,p)=h_{\tau^{\epsilon}}^{\epsilon}(q,p) \label{ftau}%
\end{equation}
and this map is close to the free-flight map $F_{\circ}^{\epsilon}$ (see
Section \ref{bmd}) of the billiard in $D^{\epsilon}$:
\begin{equation}
F_{\circ}^{\epsilon}(q,p)=b_{\tau_{\circ}^{\epsilon}-0}(q,p) \label{ftau0}%
\end{equation}
where $\tau^{\epsilon}(q,p)$ is the time the smooth Hamiltonian orbit of
$(q,p)$ needs to reach $\partial D^{\epsilon}$, and $\tau_{\circ}^{\epsilon
}(q,p)$ denotes the same for the billiard orbit. Note that we cannot claim the
closeness of the time $\tau$ maps for the smooth Hamiltonian and billiard
flows everywhere in $D^{\epsilon}$, still we claim that the maps (\ref{ftau})
and (\ref{ftau0}) are close; we will return to this later.

Outside $D^{\epsilon}$, the overall effect of the motion of smooth orbits is
close to that of a billiard reflection. Namely, as it is proved in Theorem
\ref{Main thr}, once $\nu_{i}$ is chosen such that $M_{i}^{(r)}(\nu
_{i},\epsilon)\rightarrow0$, the smooth trajectories which enter the region
$W_{i}(Q;\epsilon)\geq\nu_{i}$ at a bounded away from zero angle to the
boundary, spend in this region a small interval of time (denoted by $\tau
_{c}^{\epsilon}(q_{in},p_{in})$) after which they return to the boundary
$W_{i}(Q;\epsilon)=\nu_{i}$ (namely to $Q(q;\epsilon)=Q_{i}+\eta_{i}%
(\epsilon)$). Thus, these orbits define the map
\[
R^{\epsilon}:(q_{in}\in\partial D^{\epsilon},p_{in}%
\;\;\mbox{ looking outwards }\;D^{\epsilon})\rightarrow(q_{out}\in\partial
D^{\epsilon},p_{out}\;\;\mbox{
looking inwards }\;D^{\epsilon}).
\]

\begin{figure}[ptb]
\centering \psfig{figure =
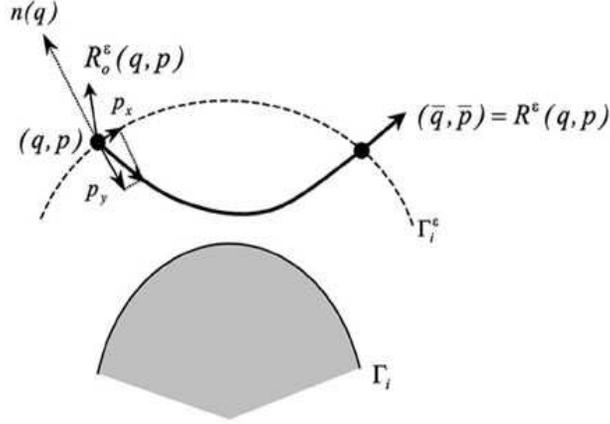,height=60mm,width=83mm}\caption{Reflection from the boundary
$\Gamma_{i}$. A smooth trajectory is marked by a bold line. An auxiliary
billiard trajectory only changes its direction according to the law
(\ref{bril}).}%
\label{fig:reflection}%
\end{figure}

\begin{figure}[ptb]
\centering \psfig{figure =
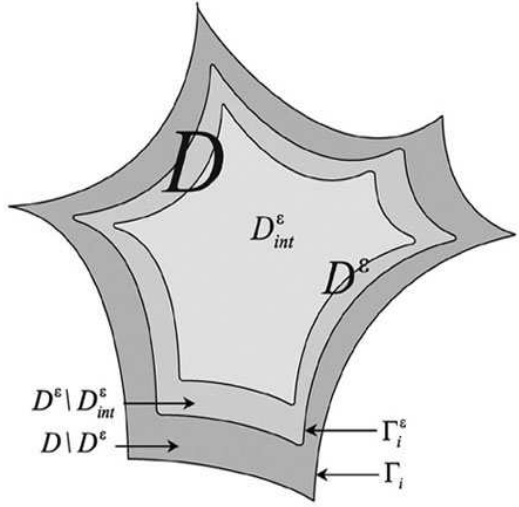,height=70mm,width=70mm}\caption{The partition of the domain
$D$ into regions: $D^{\epsilon}_{int}\subset D^{\epsilon}$. }%
\label{fig:D_structure}%
\end{figure}

It follows from the proof of Theorem \ref{Main thr} that the map $R^{\epsilon
}$ is close to the standard reflection law $R_{\circ}^{\epsilon}$ from the
boundary $\partial D^{\epsilon}$:
\begin{equation}
R_{\circ}^{\epsilon}(q,p)=\left(  q,p-2n(q)\left\langle n(q),p\right\rangle
\right)  , \label{bril}%
\end{equation}
where $n(q)$ is the unit normal vector to the boundary $\partial D^{\epsilon}$
at the point $q$. See Figure \ref{fig:reflection}. Note that the smooth
reflection law $R^{\epsilon}$ corresponds to a non-zero (though small)
collision time $\tau_{c}^{\epsilon}(q,p)$, unlike the billiard reflection
$R_{\circ}^{\epsilon}$ which happens instantaneously. Summarizing, from the
proof of Theorem \ref{Main thr} we extract that on the cross-section
\begin{equation}
S_{\epsilon}=\{\rho=(q,p):q\in\partial D^{\epsilon},\langle p,n(q)\rangle>0\}
\label{crosse}%
\end{equation}
the Poincar\'{e} map
\begin{equation}
\Phi^{\epsilon}=R^{\epsilon}\circ F^{\epsilon} \label{smbph}%
\end{equation}
is defined for the smooth Hamiltonian flow (for regular orbits - orbits which
intersect $\partial D^{\epsilon}$ at an angle bounded away from zero), and
this map is $C^{r}$-close to the billiard map $B^{\epsilon}=R_{\circ
}^{\epsilon}\circ F_{\circ}^{\epsilon}$. As the billiard map $B^{\epsilon}$ is
close to the original billiard map $B$, we obtain the closeness of the
Poincar\'{e} map $\Phi^{\epsilon}$ to $B$ as well. However, when developing
asymptotic expansions for $\Phi^{\epsilon}$, it is convenient to use the map
$B^{\epsilon}$ (rather than $B$) as the zeroth order approximation for
$\Phi^{\epsilon}$. Then, the next term in the asymptotic may be explicitly
found (see below) and the whole asymptotic expansion may be similarly developed.

We start with the estimates for the \textquotedblleft free
flight\textquotedblright\ segment of the motion, i.e. for the smooth
Hamiltonian trajectories inside $D^{\epsilon}$. For every boundary surface
$\Gamma_{i}$, choose some $\delta_{i}(\epsilon)\rightarrow0$ such that the
surfaces $Q(q;\epsilon)=Q_{i}+\delta_{i}(\epsilon)$ bound the region
$D_{int}^{\epsilon}$ inside $D^{\epsilon}$ in which the potential $V$ tends to
zero uniformly along with all its derivatives. See Figure
\ref{fig:D_structure}. Let
\begin{equation}
m^{(r)}(\delta;\epsilon)=\sup_{{\footnotesize
\begin{array}
[c]{c}%
q\in D_{int}^{\epsilon}\\
1\leq l\leq r+1
\end{array}
}}\Vert\partial^{l}V(q;\epsilon)\Vert. \label{m-small}%
\end{equation}
According to Condition I, $m$ approaches zero as $\epsilon\rightarrow0$ for
any fixed $\delta$ of the appropriate signs, therefore the same holds true for
any choice of sufficiently slowly tending to zero $\delta_{i}(\epsilon)$. As
$m^{(r)}\rightarrow0$, it follows that within $D_{int}^{\epsilon}$ the flow of
the smooth Hamiltonian trajectories is $C^{r}$-close to the free flight, i.e.
to the billiard flow. In other words, the time $\tau$ map $h_{\tau}^{\epsilon
}(q,p)=(q_{\tau},p_{\tau})$ of the smooth flow in $D_{int}^{\epsilon}$ is
$O_{_{C^{r}}}(m^{(r)})$-close to the time $\tau$ map of the billiard flow
\begin{equation}
b_{\tau}(q,p)=%
\begin{pmatrix}
q+p\tau\\
p
\end{pmatrix}
. \label{btau}%
\end{equation}
Note that on the boundary of $D^{\epsilon}$ we have, by construction,
$\mathcal{Q}_{\;i}^{\prime}(W;\epsilon)\rightarrow0$, i.e. $W_{i}^{\prime
}(Q;\epsilon)\rightarrow\infty$, while on the boundary of $D_{int}^{\epsilon}$
we have $W_{i}^{\prime}(Q;\epsilon)\rightarrow0$. Thus, we have a boundary
layer $D^{\epsilon}\backslash D_{int}^{\epsilon}$ of a non-zero width
$\left\vert \delta_{i}(\epsilon)-\eta_{i}(\epsilon)\right\vert $ in which the
gradient of the potential rapidly decreases. The speed with which the value of
$Q(q(t);\epsilon)$ changes within this boundary layer is bounded away from
zero (see the proof of Theorem \ref{Main thr}), so the time the orbit needs to
penetrate it is $O(\delta_{i})$. Within this boundary layer the time $\tau$
map $(q,p)\mapsto(q_{\tau},p_{\tau})$ of the smooth flow is not necessarily
close to the time $\tau$ map of the billiard flow (\ref{btau}). However, it is
shown in the proof of Theorem \ref{Main thr}, that the maps from one surface
$Q=const$ to any other such surface within the boundary layer are $C^{r}%
$-close for the two flows. This, obviously, implies the closeness of the maps
$F^{\epsilon}$ and $F_{\circ}^{\epsilon}$ (because the corresponding
cross-section is the surface of the kind $Q=const$ indeed).

In Appendix \ref{sec:Picard} we show that by an appropriate change of
coordinates in each of the three regions we consider (inside $D_{int}%
^{\epsilon}$, in $D^{\epsilon}\backslash D_{int}^{\epsilon}$, and outside
$D^{\epsilon}$), the equations of motion may be written as differential
equations integrated over a finite interval with a right hand side which tends
to zero in the $C^{r}$-topology as $\epsilon\rightarrow0$. Thus, not only do
we obtain error estimates for the zeroth order approximation, we also find a
method for obtaining higher order corrections using Picard iterations; The
asymptotic behavior of the right hand side of the equations leads to a
contractivity constant which asymptotically vanishes and thus the Picard
iteration scheme provides asymptotic for the solutions (each new iteration
provides a better asymptotic). In this way we prove in appendix
\ref{sec:proof of frreflights} the following

\begin{lem}
\label{freeflight} Let $q$ be an inner point of $D$, and $p$ be such that the
first hit of the billiard orbit of $(q,p)$ with the boundary $\Gamma_{i}$ is
non-tangent. Then, the orbit of the smooth flow hits the cross-section
$\{q\in\Gamma_{i}^{\epsilon}\}=\{Q(q;\epsilon)=Q_{i}+\eta_{i}(\epsilon)\}$ at
the point $(q_{\tau},p_{\tau})$ such that
\begin{equation}%
\begin{array}
[c]{ccc}%
q_{\tau} & = & q+p\tau+O_{_{C^{r}}}(m^{(r)}+\nu_{i})\\
& = & q+p\tau+\int_{0}^{\tau}\nabla V(q+ps;\epsilon)(s-\tau)ds+O_{_{C^{r-1}}%
}((m^{(r)}+\nu_{i})^{2}),\\
p_{\tau} & = & p+O_{_{C^{r}}}(m^{(r)}+\nu_{i})\\
& = & p-\int_{0}^{\tau}\nabla V(q+ps;\epsilon)ds+O_{_{C^{r-1}}}((m^{(r)}%
+\nu_{i})^{2}),
\end{array}
\label{lemfre}%
\end{equation}
where $\tau=\tau^{\epsilon}(q,p)$ denotes the travel time to the boundary of
$D^{\epsilon}$ (so $Q(q_{\tau};\epsilon)=Q_{i}+\eta_{i}(\epsilon)$):
\begin{equation}%
\begin{array}
[c]{ccc}%
\tau^{\epsilon}(q,p) & = & \tau_{\circ}^{\epsilon}(q,p)+O_{_{C^{r}}}%
(m^{(r)}+\nu_{i})\\
& = & \tau_{\circ}^{\epsilon}(q,p)+\frac{\langle\nabla Q,\int_{0}^{\tau
_{\circ}^{\epsilon}}\nabla V(q+ps;\epsilon)(\tau_{\circ}^{\epsilon
}-s)ds\rangle}{\langle\nabla Q,p\rangle}+O_{_{C^{r-1}}}((m^{(r)}+\nu_{i}%
)^{2}),
\end{array}
\label{tauint}%
\end{equation}
where $\nabla Q$ is evaluated at the (auxiliary) billiard collision point
$q+p\tau_{\circ}^{\epsilon}(p,q)$, and $\tau_{\circ}^{\epsilon}(p,q)$ is the
time the billiard orbit of $(q,p)$ needs to reach $\Gamma_{i}^{\epsilon}$.
\end{lem}

Now, let us estimate the free-flight map $F^{\epsilon}$ of the Hamiltonian
flow. If $q\in\Gamma_{j}^{\epsilon}$ and $\langle p,n(q) \rangle$ is positive
and bounded away from zero, and if the straight line issued from $q$ in the
direction of $p$ first intersects $\partial D^{\epsilon}$ (say, the surface
$\Gamma_{i}^{\epsilon}$) transversely as well (in our notations this can be
expressed as the condition that $\langle p, n(q+p\tau^{\epsilon}_{\circ
}(q,p))\rangle$ is negative and bounded away from zero), then the orbits of
the Hamiltonian flow define the map $F^{\epsilon}$ from a small neighborhood
of $(q,p)$ on the cross-section $\{q\in\Gamma_{j}^{\epsilon}\}$ in phase space
into a small neighborhood of the point $(q+p\tau^{\epsilon}_{\circ}(q,p),p)$
on the cross-section $\{q\in\Gamma_{i}^{\epsilon}\}$. See Figure
\ref{fig:free flight}. Take an inner point $(q_{1},p_{1})$ on the smooth
Hamiltonian trajectory of $(q,p)$. By construction (see (\ref{ftau})),
\[
\tau^{\epsilon}(q,p)=\tau^{\epsilon}(q_{1},-p_{1})+\tau^{\epsilon}(q_{1}%
,p_{1}),
\]
\[
(q,p)=h^{\epsilon}_{-\tau^{\epsilon}(q_{1},-p_{1})}(q_{1},p_{1})
\]
and
\[
F^{\epsilon}(q,p)=h^{\epsilon}_{\tau^{\epsilon}(q_{1},p_{1})}(q_{1},p_{1}).
\]
As $q_{1}$ is bounded away from the billiard boundary, we can plug
(\ref{lemfre}) and (\ref{tauint}) in these relations, which gives us the following

\begin{lem}
\label{fbilem} Near the point $(q,p)$ under consideration, the free flight map
$F^{\epsilon}:(q,p)\mapsto(q_{_{\tau^{\epsilon}}},p_{_{\tau^{\epsilon}}})$ for
the smooth Hamiltonian flow is $O_{_{C^{r}}}(m^{(r)}+\nu_{i}+\nu_{j})$-close
to the free flight map $F_{\circ}^{\epsilon}$ of the billiard in $D^{\epsilon
}$ and is given by
\begin{equation}%
\begin{array}
[c]{l}%
q_{_{\tau^{\epsilon}}}=q+p\tau^{\epsilon}+\int_{0}^{\tau^{\epsilon}}\nabla
V(q+ps;\epsilon)(s-\tau^{\epsilon})ds+O_{_{C^{r-1}}}((m^{(r)}+\nu_{i}+\nu
_{j})^{2}),\\
p_{_{\tau^{\epsilon}}}=p-\int_{0}^{\tau^{\epsilon}}\nabla V(q+ps;\epsilon
)ds+O_{_{C^{r-1}}}((m^{(r)}+\nu_{i}+\nu_{j})^{2}).
\end{array}
\label{fbi}%
\end{equation}
The flight time $\tau^{\epsilon}(q,p)$ is $O_{_{C^{r}}}(m^{(r)}+\nu_{i}%
+\nu_{j})$-close $\tau_{\circ}^{\epsilon}(p,q)$ and is uniquely defined by the
condition $Q(q_{\tau^{\epsilon}};\epsilon)=Q_{i}+\eta_{i}(\epsilon)$
(cf.(\ref{tauint})):
\begin{equation}
\tau^{\epsilon}(q,p)=\tau_{\circ}^{\epsilon}(q,p)+\frac{\langle\nabla
Q,\int_{0}^{\tau_{\circ}^{\epsilon}}\nabla V(q+ps;\epsilon)(\tau_{\circ
}^{\epsilon}-s)ds\rangle}{\langle\nabla Q,p\rangle}+O_{_{C^{r-1}}}%
((m^{(r)}+\nu_{i}+\nu_{j})^{2}), \label{taubo}%
\end{equation}
where $\nabla Q$ is taken at the billiard collision point $q+p\tau_{\circ
}^{\epsilon}(p,q)$ and $\tau_{\circ}^{\epsilon}(p,q)$ corresponds to the free
flight travel time: $Q(q+p\tau_{\circ}^{\epsilon}(p,q);\epsilon)=Q_{i}%
+\eta_{i}(\epsilon).$
\end{lem}

This could be written as%

\[
F^{\epsilon}=F^{\epsilon}_{\circ}+O_{_{C^{r}}}(m^{(r)}+\nu_{i}+\nu_{j})=
F^{\epsilon}_{\circ}+F^{\epsilon}_{1}+O_{_{C^{r-1}}}((m^{(r)}+\nu_{i}+\nu
_{j})^{2}),
\]
where $F^{\epsilon}_{1}=O_{_{C^{r}}}(m^{(r)}+\nu_{i}+\nu_{j})$ and
$F^{\epsilon}_{\circ}$ is defined by \ref{ftau0}.

Note that the above estimates hold true for any choice of $\delta_{i}$'s such
that $m^{(r)}\rightarrow0$. Therefore, one may take $\delta_{i}$'s tending to
zero as slow as needed in order to ensure as good estimates as possible for
the error terms in (\ref{fbi}),(\ref{taubo}).

Next we estimate the reflection law $R^{\epsilon}$ for the smooth orbit.
Consider a point $q\in\Gamma_{i}^{\epsilon}$ and let the momentum $p$ be
directed outside $D^{\epsilon}$, at a bounded from zero angle with $\Gamma
_{i}^{\epsilon}$. As we explained, the smooth trajectory of $(q,p)$ spends a
small time $\tau_{c}^{\epsilon}(q,p)$ outside $D^{\epsilon}$ and then returns
to $\Gamma_{i}^{\epsilon}$ with the momentum directed strictly inside
$D^{\epsilon}$. Let $p_{y}$ and $p_{x}$ denote the components of momentum,
respectively, normal and tangential to the boundary $\Gamma_{i}^{\epsilon}$ at
the point $q$:
\begin{equation}
p_{y}=\langle n(q),p\rangle,\qquad p_{x}=p-p_{y}n(q). \label{eq:normal-py}%
\end{equation}
We assume that the unit normal $n(q)$ is oriented inside $D^{\epsilon}$, so
$p_{y}<0$ at the initial point. Denote by $Q_{y}(q;\epsilon)$ the derivative
of $Q$ in the direction of $n(q)$:
\[
Q_{y}(q;\epsilon):=\langle\nabla Q(q;\epsilon),n(q)\rangle,
\]
let $K(q;\epsilon)$ denote the derivative of $n(q)$ in the directions tangent
to $\Gamma_{i}^{\epsilon}$, and let $l(q;\epsilon)$ denote the derivative of
$n(q)$ in the direction of $n(q)$. Obviously, $Q_{y}$ is a scalar, $K$ is a
matrix and $l$ is a vector tangent to $\Gamma_{i}^{\epsilon}$ at the point
$q$. Note that $Q_{y}\neq0$ by virtue of Condition IIc. Define the integrals:
\begin{equation}%
\begin{array}
[c]{l}%
I_{1}=I_{1}(q,p)=2\int_{0}^{-p_{y}}\mathcal{Q}_{\;i}^{\prime}(\frac
{1-p_{x}^{2}-s^{2}}{2};\epsilon)ds\\
I_{2}=I_{2}(q,p)=2\int_{0}^{-p_{y}}\mathcal{Q}_{\;i}^{\prime}(\frac
{1-p_{x}^{2}-s^{2}}{2};\epsilon)s^{2}ds,
\end{array}
\label{i12}%
\end{equation}
and the vector $J$:
\begin{equation}
J(q,p)=\left[  -\frac{I_{2}(q,p)}{p_{y}}l(q;\epsilon)+I_{1}(q,p)K(q;\epsilon
)p_{x}\right]  /Q_{y}(q;\epsilon). \label{jvec}%
\end{equation}
Notice that $J$ is a vector tangent to $\Gamma_{i}^{\epsilon}$ at the point
$q$ and that by (\ref{M-big}),
\begin{equation}
I_{1,2}=O_{_{C^{r}}}(M_{i}^{(r)}),J=O_{_{C^{r-1}}}(M_{i}^{(r)}). \label{i12es}%
\end{equation}

In Appendix \ref{sec:proof of frreflights} we prove the following

\begin{lem}
\label{error_of_reflection} For the smooth Hamiltonian flow, the collision
time is estimated as
\begin{equation}
\tau_{c}^{\epsilon}(q,p)=O_{_{C^{r}}}(M_{i}^{(r)})=-\frac{1}{Q_{y}%
(q;\epsilon)}I_{1}(q,p)+O_{_{C^{r-1}}}((M_{i}^{(r)})^{2}). \label{taucsh}%
\end{equation}
The reflection map $R^{\epsilon}:(q,p)\mapsto(\bar{q},\bar{p})$ is given by:
\begin{equation}%
\begin{array}
[c]{l}%
\bar{q}=q+O_{_{C^{r}}}(M_{i}^{(r)})=q+p_{x}\tau_{c}^{\epsilon}%
(q,p)+O_{_{C^{r-1}}}((M_{i}^{(r)})^{2}),\\
\bar{p}=p-2n(q)p_{y}+O_{_{C^{r}}}(M_{i}^{(r)})=p-2n(q)p_{y}-p_{y}%
J(q,p)-n(q)\langle p_{x},J(q,p)\rangle+O_{_{C^{r-1}}}((M_{i}^{(r)})^{2}).
\end{array}
\label{interm}%
\end{equation}

\end{lem}

As we see from this lemma (see also (\ref{i12es})),
\[
R^{\epsilon}=R_{\circ}^{\epsilon}+O_{_{C^{r}}}(M_{i}^{(r)})=R_{\circ
}^{\epsilon}+R_{1}^{\epsilon}+O_{_{C^{r-1}}}((M_{i}^{(r)})^{2}),
\]
where $R_{1}^{\epsilon}=O_{_{C^{r-1}}}(M_{i}^{(r)})$ and $R_{\circ}^{\epsilon
}$ is defined by \ref{bril}. Thus, the smooth reflection law is $O_{_{C^{r}}%
}(M_{i}^{(r)})$-close to the billiard reflection law (\ref{bril}).

\begin{figure}[ptb]
\centering \psfig{figure =
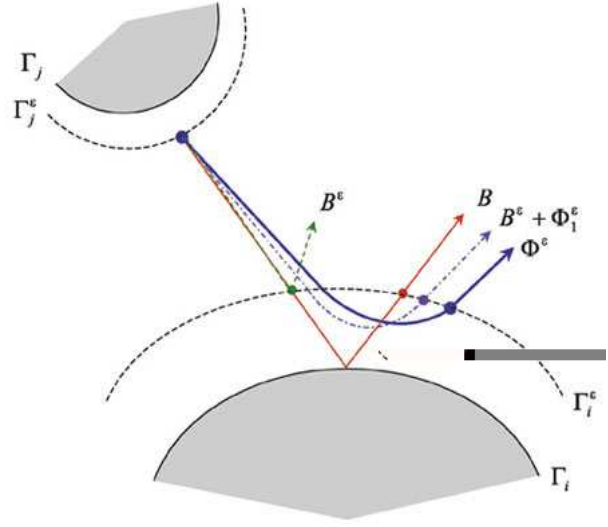,height=70mm,width=80mm}\caption{Billiard map $B$ (red).
Billiard map of the auxiliary billiard $B^{\epsilon}$ (green). Poincar\'{e}
map for the smooth Hamiltonian flow $\Phi^{\epsilon}$ (blue). The first
approximation of $\Phi^{\epsilon}$ $B^{\epsilon}+\Phi_{1}^{\epsilon}$
(violet).}%
\label{fig:Poincare_smooth}%
\end{figure}Combining the above lemmas we establish:

\begin{theorem}
\label{thm-est} Let the potential $V(q;\epsilon)$ satisfy Conditions
\textbf{I-IV}, and choose $\delta_{i}$'s and $\nu_{i}$'s such that $\delta
_{i}(\epsilon),\nu_{i}(\epsilon),m^{(r)}(\epsilon),M_{i}^{(r)}(\epsilon
)\rightarrow0$ as $\epsilon\rightarrow0$. Then, on the cross-section
$S_{\epsilon}$ (see (\ref{crosse})) near orbits of a regular
reflection\footnote{that is, given any constant $C>0$, near the points
$(q,p)\in S_{\epsilon}$ such that $\langle n(q),p\rangle\geq C$ and
$\left\vert \langle n(\bar{q}),\bar{p}\rangle\right\vert \geq C$ where
$(\bar{q},\bar{p})=B^{\epsilon}(q,p)$}, for all sufficiently small $\epsilon$
the Poincar\'{e} map $\Phi^{\epsilon}$ of the smooth Hamiltonian flow is
defined, and it is $O(m^{(r)}+\nu+M^{(r)})$-close in the $C^{r}$-topology to
the billiard map $B^{\epsilon}=R_{\circ}^{\epsilon}\circ F_{\circ}^{\epsilon}$
in the auxiliary billiard table $D^{\epsilon}$ (see Figure
\ref{fig:Poincare_smooth}). Furthermore,
\begin{multline}
\Phi^{\epsilon}=R^{\epsilon}\circ F^{\epsilon}=B^{\epsilon}+O_{_{C^{r}}%
}(m^{(r)}+\nu+M^{(r)})=(R_{\circ}^{\epsilon}+R_{1}^{\epsilon})\circ(F_{\circ
}^{\epsilon}+F_{1}^{\epsilon})+O_{_{C^{r-1}}}((m^{(r)}+\nu+M^{(r)}%
)^{2})\label{one reflection error}\\
=:B^{\epsilon}+\Phi_{1}^{\epsilon}+O_{_{C^{r-1}}}((m^{(r)}+\nu+M^{(r)})^{2})
\end{multline}
(where $\nu=\max_{i}\nu_{i}$, $M^{(r)}=\max_{i}M_{i}^{(r)}$,$\Phi
_{1}^{\epsilon}=O_{_{C^{r-1}}}(m^{(r)}+\nu+M^{(r)})$, and the first order
corrections $F_{1}^{\epsilon}$ and $R_{1}^{\epsilon}$ are explicitly
calculated in Lemmas \ref{fbilem} and \ref{error_of_reflection}).
\end{theorem}

\begin{theorem}
\label{thim} Given a finite $T$ and a regular billiard trajectory in $[0,T]$,
the time $t$ map of the smooth Hamiltonian flow and of the corresponding
auxiliary billiard are $O(\nu+m^{(r)}+M^{(r)})$-close in the $C^{r}$-topology
for all $t\in T\backslash T_{R}$, where $T_{R}$ is the finite collection of
impact intervals each of them of length $O(|\delta|+M^{(r)})$.
\end{theorem}

\subsubsection{Error estimates for some model potentials}

Now we can estimate the deviation of the smooth Hamiltonian trajectories from
the regular (non-tangent, non-corner) billiard ones for various concrete
potentials $V(q;\epsilon)$. To make a general estimate possible, we have to
assume that the behavior of the potential near the boundary dominates the
estimate; We say that $V(q;\epsilon)$ is \emph{boundary dominated}, if
$V(q;\epsilon)$ and its derivatives are smaller in the interior of
$D_{int}^{\epsilon}$ (i.e. in the region bounded by the surfaces
$Q(q;\epsilon)=Q_{i}+\delta_{i}(\epsilon)$) than on the boundary of this
domain. This means that for boundary dominated potentials
\begin{equation}
m^{(r)}(\delta;\epsilon)=\sup_{{\footnotesize
\begin{array}
[c]{c}%
q\in D_{int}^{\epsilon}\\
1\leq l\leq r+1
\end{array}
}}\Vert\partial^{l}V(q;\epsilon)\Vert=\sup_{{\footnotesize
\begin{array}
[c]{c}%
q\in\partial D_{int}^{\epsilon}\\
1\leq l\leq r+1
\end{array}
}}\Vert\partial^{l}V(q;\epsilon)\Vert. \label{assumption
on m}%
\end{equation}
By the definition of the pattern function $Q$, near a given boundary
$\Gamma_{i}$
\[
V(q;\epsilon)\bigg|_{q\in\partial D_{int}^{\epsilon}}\equiv W_{i}%
(Q(q;\epsilon)-Q_{i};\epsilon)\bigg|_{Q=Q_{i}+\delta_{i}}=W_{i}(\delta
_{i};\epsilon)
\]
Since $Q(q;\epsilon)$ is bounded with its derivatives, we conclude that there
exists a constant $C$ such that%
\begin{equation}
m^{(r)}(\delta;\epsilon)=C\max_{i}\max_{{\footnotesize 1\leq l\leq r+1}}%
|W_{i}^{(l)}(\delta_{i};\epsilon)|. \label{boundary-m}%
\end{equation}
Thus, for boundary dominated potentials, one can estimate the differences
$h_{t}^{\epsilon}-b_{t}$ and $\Phi^{\epsilon}-B^{\epsilon}$ in terms of the
barrier functions alone.

The corresponding estimates given by Theorems \ref{thim} and \ref{thm-est}
hold true for every choice of $\nu$ and $\delta$ such that $\delta
(\epsilon),\nu(\epsilon),m_{r}(\delta(\epsilon);\epsilon),M_{r}(\nu
(\epsilon);\epsilon)\rightarrow0$ as $\epsilon\rightarrow0$ (for simplicity of
notation we assume hereafter that the barrier function $W$ is the same for all
boundary surfaces $\Gamma_{i}$, and thus suppress the dependence on $i$). To
obtain the best estimates, we have to find $\nu(\epsilon)$ and $\delta
(\epsilon)$ which minimize the expression $\nu+M^{(r)}(\nu;\epsilon
)+m^{(r)}(\delta;\epsilon)$. In this way, we first find $\nu(\epsilon)$ which
minimizes $\nu+M^{(r)}(\nu;\epsilon)$. As $M^{(r)}$ is a decreasing function
of $\nu$ (see (\ref{M-big})), the sought $\nu(\epsilon)$ solves the equation
\begin{equation}
\nu=M^{(r)}(\nu;\epsilon). \label{eqnu}%
\end{equation}
After $\nu$ is determined, we may try to make $\delta(\epsilon)$ go to zero so
slow that the corresponding value of $m^{(r)}$ (see (\ref{boundary-m})) will
be asymptotically equal to $\nu(\epsilon)$. Once succeeded, we may conclude
that $\nu(\epsilon)$ given by (\ref{eqnu}) estimates the deviation between
regular billiard and smooth trajectories. Notice that the significance of
$\nu(\epsilon)$ is three-folded; First, it determines the optimal auxiliary
billiard which supplies the best approximation to the smooth Hamiltonian flow
(see Lemma \ref{error_of_reflection}). Second, it estimates the accuracy of
this approximation. Third, it determines, via the relation $m^{(r)}%
(\delta)=\nu$, the width $|\delta(\epsilon)|+\nu(\epsilon)$ of the boundary
layer in which the billiard and the Hamiltonian flows are not close (Theorem
\ref{thim}). Let us proceed to examples.

\begin{prop}
\label{exponential} Consider the boundary dominated potential $V(q;\epsilon)$
corresponding to the barrier function $W(Q)=e^{-\frac{Q}{\epsilon}}$ for small
$Q$. Then, near regular billiard trajectories, the smooth Hamiltonian flow is
$O(\sqrt[r+2]{\epsilon})$-close in the $C^{r}$-topology to the billiard flow
within the auxiliary billiard defined by the level set $Q(q;\epsilon
)=\eta(\epsilon)=O(\epsilon\ln\epsilon)$. The corresponding Poincar\'{e} map
$\Phi^{\epsilon}$ is $O_{_{C^{r}}}(\sqrt[r+2]{\epsilon})$-close to the
auxiliary billiard map $B^{\epsilon}$. The impact intervals lengths are
$O(\sqrt[r+2]{\epsilon})$.
\end{prop}

\begin{proof}
Since
$W^{(l)}(Q;\epsilon)=(-\epsilon)^{-l}e^{-\frac{Q}{\epsilon}}$, we
obtain that
$m^{(r)}(\delta;\epsilon)=O(\epsilon^{-(r+1)}e^{-\frac{\delta}{\epsilon}})$
(since the potential is boundary dominated, we may use
(\ref{boundary-m})). The inverse to $W(Q;\epsilon)$ is given by
$\mathcal{Q}(W;\epsilon)=-\epsilon\ln W$, so
$\mathcal{Q}^{(l)}(W;\epsilon)=(-1)^{l}(l-1)!\epsilon W^{-l}$, and
$M^{(r)} (\nu,\epsilon)=O(\epsilon\nu^{-{(r+1)}})$ (see
(\ref{M-big})). Plugging this in (\ref{eqnu}), we find
\begin{equation}
\nu(\epsilon)=\sqrt[r+2]{\epsilon}.\label{nueex}%
\end{equation}
By choosing
$\delta(\epsilon)=-(r+1+\frac{1}{r+2})\epsilon\ln\epsilon$, we
obtain $m^{(r)}(\delta,\epsilon)\sim\nu(\epsilon)$, so for $\nu$
given by (\ref{nueex}) we have that $\nu+M^{(r)}+m^{(r)}=O(\nu)$,
and the proposition now follows immediately from Theorems
\ref{thm-est} and \ref{thim} (the value of
$\eta(\epsilon)=O(\epsilon\ln\epsilon)$ is given by $\eta
=\mathcal{Q}(\nu;\epsilon)$).
\end{proof}

\begin{prop}
\label{gaussian} Let the boundary dominated potential $V(q;\epsilon)$
correspond to the barrier function $W(Q)=e^{-\frac{Q^{2}}{\epsilon}}$ for
small $Q$. Then, near the regular billiard trajectories, the smooth
Hamiltonian flow is $O(\nu(\epsilon))=O(\sqrt[2(r+2)]{\frac{\epsilon}%
{|\ln\epsilon|}})$-close in the $C^{r}$-topology to the billiard flow within
the auxiliary billiard defined by the level set $Q(q;\epsilon)=\eta
(\epsilon)=O(\sqrt{\epsilon|\ln\epsilon|})$. The corresponding Poincar\'{e}
map $\Phi^{\epsilon}$ is $O_{_{C^{r}}}(\nu(\epsilon))$-close to the auxiliary
billiard map $B^{\epsilon}$. The impact intervals are of the length
$O(\nu(\epsilon))$.
\end{prop}

\begin{proof}
It is easy to see that $W^{(l)}(Q;\epsilon)=O((\frac{Q}{\epsilon})^{l}e^{-\frac{Q^{2}%
}{\epsilon}})$ for $Q\gg\sqrt{\epsilon}$, hence
$m^{(r)}(\delta;\epsilon)=
O((\frac{\delta}{\epsilon})^{r+1}e^{-\frac{\delta^{2}}{\epsilon}})$.
From $\mathcal{Q}(W;\epsilon)=\sqrt{-\epsilon\ln W}$ we obtain
$M^{(r)}(\nu;\epsilon )=O(\sqrt{\frac{\epsilon}{|\ln\nu|}}
\nu^{-(r+1)})$. Plugging this in (\ref{eqnu}), we indeed find
\[
\nu(\epsilon) =M^{(r)}(\nu;\epsilon)=O(\sqrt[2(r+2)]{\frac{\epsilon}%
{|\ln\epsilon|}}),
\]
as required. By choosing
$\delta(\epsilon)\sim\sqrt{-\frac{1}{2}(r+1+\frac
{1}{r+2})\epsilon\ln\epsilon}$, we obtain
$m^{(r)}(\delta;\epsilon)\sim \nu(\epsilon)$, so the rest follows
directly from Theorems \ref{thm-est} and \ref{thim}.
\end{proof}

\begin{prop}
\label{coulomb} Let the boundary dominated potential $V(q;\epsilon)$
correspond to the barrier function $W(Q)=(\frac{\epsilon}{Q})^{\alpha}$. Then,
near the regular billiard trajectories, the smooth Hamiltonian flow is
$O(\nu(\epsilon))=O(\sqrt[r+2+\frac{1}{\alpha}]{\epsilon})$-close in the
$C^{r}$-topology to the billiard flow within the auxiliary billiard defined by
the level set $Q(q;\epsilon)=\eta(\epsilon)=O(\nu^{r+2})$. The corresponding
Poincar\'{e} map $\Phi^{\epsilon}$ is $O_{_{C^{r}}}(\nu(\epsilon))$-close to
the auxiliary billiard map $B^{\epsilon}$. The impact intervals are
$O(\nu(\epsilon))$ when $\alpha\geq1$, and $O(\nu(\epsilon)^{\frac
{\alpha(r+2)}{\alpha+r+1}})$ when $\alpha\leq1$.
\end{prop}

\begin{proof}
As above, using
$W^{(l)}(Q;\epsilon)=O(\frac{\epsilon^{\alpha}}{Q^{l+\alpha}})$ we
obtain that $m^{(r)}(\delta;\epsilon)=O(\frac{\epsilon^{\alpha}}%
{\delta^{r+1+\alpha}})$, and since $\mathcal{Q}(W;\epsilon)=\frac{\epsilon}%
{W^{1/\alpha}}$, we find $\mathcal{Q}^{(l)}(W;\epsilon)=O(\frac{\epsilon}%
{W^{l+1/\alpha}})$ and thus $M^{(r)}(\nu;\epsilon)=O(\frac{\epsilon}%
{\nu^{r+1+1/\alpha}}).$ It follows that
$\nu(\epsilon)=O(\sqrt[r+2+\frac {1}{\alpha}]{\epsilon})$ solves
$\nu=M^{(r)}(\nu;\epsilon)$. Now
$\eta(\epsilon)=\mathcal{Q}(\nu)=\frac{\epsilon}{\nu^{1/\alpha}}=O(\nu^{r+2}%
)$. By taking $\delta(\epsilon)=\nu^{\frac{\alpha(r+2)}{r+1+\alpha}}$, we
ensure that $m^{(r)}(\delta,\epsilon)\sim\nu(\epsilon)$. The length of impact
intervals is now given by $O(\nu+\delta)$.
\end{proof}

Note that the asymptotic for the deviation of the smooth trajectories from the
billiard ones and for the length of the impact intervals depend strongly on
$r$, i.e. on the number of derivatives (with respect to initial conditions)
which we want to control.

\subsection{\label{sec:persis}Persistence of periodic and homoclinic orbits}

The closeness of the billiard and smooth flows after one reflection leads,
using standard results, to persistence of regular periodic and homoclinic
orbits. For completeness we state these results explicitly: \label{sec:persis}

\begin{theorem}
\label{thm:persistance} Consider a Hamiltonian system with a potential
$V(q,\epsilon)$ satisfying Condition I-IV in a billiard table $D$. Let
$P^{b}(t)$ denote a non-parabolic, non-singular periodic orbit of a period $T$
for the billiard flow. Then, for any choice of $\nu(\epsilon),\delta
(\epsilon)$ such that $\nu(\epsilon),\delta(\epsilon),m^{(1)}(\epsilon
),M^{(1)}(\epsilon)\rightarrow0$ as $\epsilon\rightarrow0$, the smooth
Hamiltonian flow has a uniquely defined periodic orbit $P^{\epsilon}(t)$ of
period $T^{\epsilon}=T+O(\nu+m^{(1)}+M^{(1)})$, which stays $O(\nu
+m^{(1)}+M^{(1)})$-close to $P^{b}$ for all $t$ outside of collision intervals
(finitely many of them in a period) of length $O(|\delta|+M^{(1)})$. Away from
the collision intervals, the local Poincar\'{e} map near $P^{\epsilon}$ is
$O_{_{C^{r}}}(\nu+m^{(r)}+M^{(r)})$-close to the local Poincar\'{e} map near
$P^{b}$. In particular, if $P^{b}$ is hyperbolic, then $P^{\epsilon}$ is also
hyperbolic and, inside $D^{\epsilon}$, the stable and unstable manifolds of
$P^{\epsilon}$ approximate $O_{_{C^{r}}}(\nu+m^{(r)}+M^{(r)})$-closely the
stable and unstable manifolds of $P^{b}$ on any compact, forward-invariant or,
respectively, backward-invariant piece bounded away from the singularity set
in the billiard's phase space; furthermore, any transverse regular homoclinic
orbit to $P^{b}$ is, for sufficiently small $\epsilon$, inherited by
$P^{\epsilon}$ as well.
\end{theorem}

\begin{figure}[ptb]
\centering
\psfig{figure = 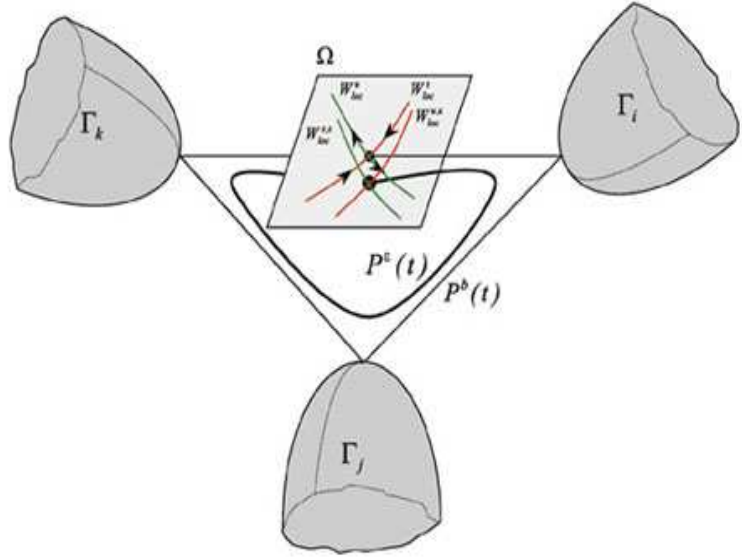,height=80mm,width=100mm}\caption{$P^{b}$ is
a billiard periodic orbit (solid). $P^{\epsilon}$ is a periodic orbit of the
smooth Hamiltonian flow (bold).}%
\end{figure}

As $P^{b}$ is a regular periodic orbit, i.e. it makes only regular reflections
from the boundary (a finite number of them on the period), it follows from
Theorem \ref{thm-est} that a Poincar\'{e} map for the smooth Hamiltonian flow
near $P^{b}$ is $O(\nu+m^{(1)}+M^{(1)})$-close in $C^{1}$ topology to the
Poincar\'{e} map of the auxiliary billiard $D^{\epsilon}$, while the latter is
$O(\eta(\epsilon))$-close to the Poincar\'{e} map for the original billiard
$D$. Moreover, from (\ref{M-big}) it follows that $\eta(\epsilon)\leq
M^{(0)}\leq M^{(1)}$ and we can conclude that a Poincar\'{e} map for the
smooth Hamiltonian flow near $P^{b}$ is $O(\nu+m^{(1)}+M^{(1)})$-close in
$C^{1}$ topology to the Poincar\'{e} map for the original billiard $D$. Since,
by assumption, $P^{b}(t)$ is non-parabolic, the corresponding fixed point of
the Poincar\'{e} map persists for sufficiently small $\epsilon$ in virtue of
the implicit function theorem (the closeness of the corresponding
continuous-time orbits is given by Theorem \ref{thim}). The continuous
dependence of the invariant manifolds of $\epsilon$ in the hyperbolic case
follows from the continuous dependence of the Poincar\'{e} map $\Phi
^{\epsilon}$ on $\epsilon$ at all $\epsilon\geq0$ (Theorem \ref{thm-est}), and
implies the persistence of transverse homoclinics immediately. Indeed, the
formulation regarding the closeness of compact pieces of the \emph{global}
stable and unstable manifolds may be easily verified by applying finite time
extensions of the local stable and unstable manifolds. Note that similar
persistence result holds true for topologically transverse homoclinic orbits.

More generally, one may claim (by the shadowing lemma) the persistence of
compact uniformly hyperbolic sets composed of regular billiard orbits. Note
that the accuracy of the approximation of smooth orbits (periodic and
aperiodic) by the billiard ones, does not depend on the orbit (e.g. is
independent of its period) and is given by the maximal deviation for each
reflection (times a constant). This holds true for any compact set of regular
orbits of a strictly dispersing billiard flow (since such billiards are
uniformly hyperbolic); see for example a nice application by Chen
\cite{Chen04}.

In some cases, to establish the existence of transverse or topologically
transverse homoclinic orbits in a family of billiard flows $b_{t}(\gamma)$ in
$D_{\gamma}$, one uses higher dimensional generalizations of the
Poincare-Melnikov integral (see Section \ref{sec:ellips}). In particular, with
the near integrable setting, the "splitting distance" between the manifolds
near the transverse homoclinic orbit may be proportional to an unfolding
parameter $\gamma$. The above theorem implies that if $\epsilon_{0}%
=\epsilon_{0}(\gamma)$ is chosen so that $\nu(\epsilon_{0},\gamma
)+m^{(1)}(\delta(\epsilon_{0},\gamma);\epsilon_{0},\gamma)+M^{(1)}%
(\nu(\epsilon_{0},\gamma);\epsilon_{0},\gamma))=o(\gamma)$ and $\epsilon
_{0}(\gamma)\rightarrow0$ as $\gamma\rightarrow0$ then, for sufficiently small
$\gamma$, transverse homoclinic orbits appear in the smooth flow for all
$\epsilon\in(0,\epsilon_{0}(\gamma))$. In the next section we use this remark
and \cite{DFRR01} to establish that transverse homoclinic orbits appear in
families of smooth billiard potentials which limit to the ellipsoidal billiard.

\section{\label{sec:ellips}Application to ellipsoidal billiards with
potential}

Consider the billiard motion in an ellipsoid%

\begin{equation}
D=\{q\in\mathbb{R}^{n}:\langle q,A^{-2}q\rangle\leq1\},
\label{generic ellipsoid}%
\end{equation}
\[
A=\Diag(d_{1},\ldots,d_{n})\;\;\;\;\;\;\;d_{1}\geq\ldots\geq d_{n}\geq0.
\]
The ellipsoid is called generic if all the above inequalities are strict. A
well known result of Birkhoff \cite{Birk27} is that the billiard motion in an
ellipsoid is integrable, and the mathematical theory which may be invoked to
describe and generalize this result is still under development - see Radnovic
\cite{DrR04} and references therein. Delshams \textit{et al} \cite{DFRR01} and
recently Bolotin \textit{et al} \cite{BDRR04} (see also references therein)
investigate when small non-quadratic symmetric perturbations to the
ellipsoidal shape change the integrability property. In this series of works
the authors prove the persistence of some symmetric homoclinic orbits, and for
specific cases they prove that these orbits are transverse homoclinic orbits
of the perturbed billiard, thus proving that integrability is destroyed. Here,
we show that using the machinery we developed we can immediately extend their
work to the smooth billiard-potential case (notice that in \cite{BDRR04} some
results are extended to billiards with a $C^{2}$-small Hamiltonian
perturbation in the domain's interior, however the billiard potentials which
we consider do not fall into this category - near the boundary they correspond
to a large perturbation even in the $C^{1}$-norm). We will first explain the
relevant main results of Delshams et al, then supply the corresponding
proposition for the smooth case (consequences of Theorem \ref{Main thr}, or
more specifically of Theorem \ref{thm:persistance}) and then the corresponding
quantitative estimates for specific potentials (which follows from
Propositions \ref{exponential}-\ref{coulomb}).

\subsection{The billiard in a perturbed ellipsoid}

Consider the simplest unstable periodic orbit in an ellipsoidal billiard - the
orbit along the diameter of the ellipsoid joining the vertices $(-d_{1}%
,0,\ldots,0)$ and $(d_{1},0,\ldots,0).$ Denote the set formed by the
two-periodic points associated with the diameter by%
\begin{equation}
P^{b}=\{\rho_{+},\rho_{-}\}\quad\rho_{\pm}=\{q_{\pm},p_{\mp}\}\quad q_{\pm
}=(\pm d_{1},0,\ldots,0)\quad p_{\pm}=(\pm1,0,\ldots,0).
\label{2-periodic points}%
\end{equation}
These points correspond to isolated two-periodic hyperbolic orbits of the
Billiard map $B$ and the corresponding periodic orbit $P_{t}^{b}=b_{t}%
(\rho_{+})$ of the billiard flow. The $n-1$-dimensional ($n$-dimensional for
the flow) stable and unstable manifolds of this periodic orbit coincide; In
$2$-dimensions there are 4 separatrices connecting $\{\rho_{+},\rho_{-}\}$
whereas the topology of the separatrices in the higher dimensional case is
non-trivial - it is well described by CW complexes for the 3 dimensional case
and by hierarchal structure of separatrix submanifolds in the higher
dimensional case (see \cite{DFRR01}).

Of specific interest are the symmetric homoclinic orbits - it is established
in \cite{DFRR01} that in the generic $2$ dimensional case there are exactly
$4$ homoclinic orbits which are $x-$symmetric (symmetric, in the configuration
space, to reflections about the $x$-axis) and $4$ which are $y-$symmetric. In
the generic $3$ dimensional case, in addition to the $16$ planar symmetric
orbits ($8$ in each of the symmetry planes- $xy$ and $xz$) there are $16$
additional symmetric spatial orbits - $8$ are symmetric with respect to
reflection about the $xz$ plane and $8$ are $y$ axial. In the $n$ dimensional
case there are $2^{n+1}$ spatial symmetric orbits.

\begin{figure}[ptb]
\centering \psfig{figure =
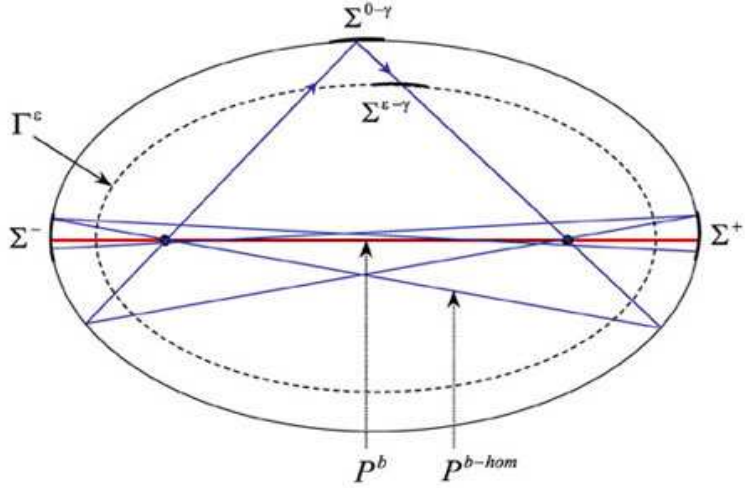,height=70mm,width=100mm}\caption{Billiard trajectory
giving rise to a y-symmetric homoclinic orbit (blue).}%
\label{fig:homoclinic_orbit}%
\end{figure}

Denote by $P^{b-\hom}=\left\{  P_{i}^{b-\hom}\right\}  _{i=-\infty}^{\infty}$
one of these symmetric homoclinic orbits of the billiard map in the ellipsoid,
so $P_{i+1}^{b-\hom}=BP_{i}^{b-\hom}$ and $P_{t}^{b-\hom}=b_{t}(P_{0}^{b-\hom
})$ denotes the corresponding continuous orbit of the billiard flow. Given a
$\varsigma$ such that $0<\varsigma\ll d_{n}$, define the local cross-sections
of the billiard map by:%
\begin{align*}
\Sigma^{-}  &  =\{(q,p)|q\in\partial D,q_{1}+d_{1}<\varsigma,\text{ }%
1-p_{1}<\varsigma\},\\
\Sigma^{+}  &  =\{(q,p)|q\in\partial D,d_{1}-q_{1}<\varsigma,\text{ }%
p_{1}+1<\varsigma\},
\end{align*}
so, in particular, $\rho_{\pm}\in$ $\Sigma^{\pm}$ and $\Sigma^{\pm}\subset S$,
where $S$ is the natural cross-section on which the billiard map is defined
(see Section \ref{bmd}). It follows that only a finite number of points in
$P^{b-\hom}$ do not fall into $\Sigma^{\pm}$, and that for any given geometry
there exist a finite $\varsigma$ such that $P^{b-\hom}\backslash\{P^{b-\hom
}\cap\Sigma^{\pm}\}\neq\varnothing$ for all the symmetric orbits. See Figure
\ref{fig:homoclinic_orbit}. Thus, it is possible to choose $P_{0}^{b-\hom}$
and a local cross-section $\Sigma^{0}$ such that $P_{0}^{b-\hom}\in\Sigma
^{0}\subset\{S\backslash\{\Sigma^{+}\cup\Sigma^{-}\}\}$. Notice that for the
ellipsoid all the reflections are regular, and furthermore, for the symmetric
homoclinic orbits, if $d_{1}$ is finite and $d_{n}$ is positive then all the
reflection angles of $P^{b-\hom}$ are strictly bounded away from $\pi/2$.

Now, consider a \textbf{symmetric }perturbation of the ellipsoid $Q$ of the form:%

\begin{equation}
D_{\gamma}=\{q\in\mathbb{R}^{n}:\langle q,A^{-2}q\rangle\leq1+\gamma\Xi
(\frac{q_{1}^{2}}{d_{1}^{2}},\ldots,\frac{q_{n}^{2}}{d_{n}^{2}})\},
\label{geomer perturb ellipsoid}%
\end{equation}
where the hypersurface $D_{\gamma}\in\mathbb{R}^{n+1}$ is symmetric with
regard to all the coordinate axis of $\mathbb{R}^{n}$ and the function
$\Xi:\mathbb{R}^{n}\rightarrow\mathbb{R}$ is either a general \emph{entire}
function, such that $\Xi(0,\ldots,0)=0$ or of a specific form (e.g.
quadratic). By using symmetry arguments, Delshams \textit{et al} \cite{DFRR01}
prove that for generic billiard the above mentioned symmetric homoclinic
orbits persist under such symmetric perturbations. Furthermore, analyzing the
asymptotic properties of the symplectic discrete version of the
Poincar\'{e}-Melnikov potential (the high dimensional analog of the integral),
they prove that for sufficiently small perturbations (small $\gamma$) the
$n$-dimensional symmetric homoclinic orbits are transverse in the following
four cases:

\begin{enumerate}
\item In two-dimensions, for narrow ellipses ($\beta_{1}=\frac{d_{2}^{2}%
}{d_{1}^{2}}\ll1$), for any analytic small enough symmetric perturbation.

\item In two-dimensions, in the non-circular case ($\beta_{1}\neq1$), for
$\Xi(\frac{x^{2}}{d_{1}^{2}},\frac{y^{2}}{d_{2}^{2}})=\frac{y^{4}}{d_{2}^{4}}$.

\item In the three-dimensional case, for nearly flat ellipses ($\beta
_{2}=\frac{d_{3}^{2}}{d_{1}^{2}}\ll1$), for perturbations of the form:
$\Xi(\frac{x^{2}}{d_{1}^{2}},\frac{y^{2}}{d_{2}^{2}},\frac{z^{2}}{d_{3}^{2}%
})=\frac{z^{2}}{d_{3}^{2}}R(\frac{y^{2}}{d_{2}^{2}},\frac{z^{2}}{d_{3}^{2}})$
where $R$ is a generic polynomial (or of some specific list).

\item In the three-dimensional case, for nearly oblate ellipses ($\beta
_{1}=\frac{d_{2}^{2}}{d_{1}^{2}}\simeq1$), for the perturbation $\Xi
(\frac{x^{2}}{d_{1}^{2}},\frac{y^{2}}{d_{2}^{2}},\frac{z^{2}}{d_{3}^{2}%
})=\frac{z^{2}}{d_{3}^{2}}\frac{y^{2}}{d_{2}^{2}}$.
\end{enumerate}

To establish these results, the Poincar\'{e}-Melnikov potential is calculated
for each of these cases, and it is shown that it has non-degenerate critical
points at the corresponding symmetric trajectories. It follows that
$P^{b-\hom-\gamma}$ persists and the change in the splitting distance between
the separatrices $\mathcal{W}^{u}$ and $\mathcal{W}^{s}$ near $P_{0}%
^{b-\hom-\gamma}$ is proportional to $\gamma$, the perturbation amplitude, so
that near $P_{0}^{b-\hom-\gamma}$ at the local cross-section $\Sigma
^{0-\gamma}$,%

\begin{equation}
d(W_{\gamma}^{s},W_{\gamma}^{u})=M(r)\gamma+O(\gamma^{2}).
\label{perturbation of geometry}%
\end{equation}
where $r\in R^{n-1}$ denotes some parametrization along $\mathcal{W}$ and
$M(r)$ (the gradient of the Poincar\'{e}-Melnikov potential) has simple zeroes
at the parameter values corresponding to any of the spatial symmetric
homoclinic orbits $P_{0}^{b-\hom}$.

\subsection{Smooth Potential in a near ellipsoidal region}

\begin{figure}[ptb]
\centering \psfig{figure =
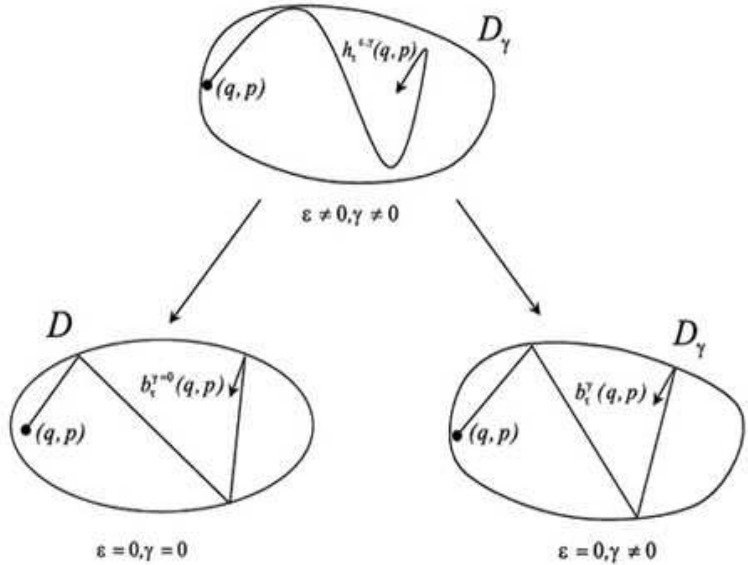,height=80mm,width=100mm}\caption{Perturbation of a billiard flow
inside a perturbed ellipsoid family $D_{\gamma}$.}%
\label{fig:ellipsoid}%
\end{figure}

Let us now consider a two parameter family of smooth potentials $V(q;\gamma
,\epsilon)$ which limit, as $\epsilon\rightarrow0$ to the billiard flow in the
perturbed ellipsoid family $D_{\gamma}$; namely, consider the family of
Hamiltonian flows:%

\begin{equation}
H(\epsilon,\gamma)=\frac{p^{2}}{2}+V(q;\gamma,\epsilon).
\label{eq:hambilellipse}%
\end{equation}
where $V(q;\gamma,\epsilon)$ satisfies conditions I-IV for all $\gamma$
values. In the four cases mentioned above, the flow limits, as $\epsilon
\rightarrow0,$ to an integrable billiard motion inside the ellipsoid $D$ when
$\gamma=0$ and, for $\gamma\neq0,$ to a non-integrable billiard motion inside
the perturbed ellipsoid $D_{\gamma}$. See Figure \ref{fig:ellipsoid}.

Applying Theorem \ref{thm:persistance} to an interior transverse local return
map near $\Sigma^{0-\gamma}$, and noticing that all homoclinic orbits of the
billiard flow in $D_{\gamma}$ are regular orbits, we immediately establish:

\begin{prop}
\label{prop:ellip-pers}Consider the Hamiltonian flow (\ref{eq:hambilellipse}),
where $V(q;\gamma,\epsilon)$ is a billiard potential limiting to the billiard
flow in $D_{\gamma}$ ($V(q;\gamma,\epsilon)$ satisfies conditions I-IV for all
$\gamma$ values). Let the function $\epsilon_{0}(\gamma)$ satisfy%
\[
\nu(\epsilon_{0},\gamma)+m^{(1)}(\delta(\epsilon_{0},\gamma);\epsilon
_{0},\gamma)+M^{(1)}(\nu(\epsilon_{0},\gamma);\epsilon_{0},\gamma))=o(\gamma)
\]
and $\epsilon_{0}(\gamma)$ $\rightarrow0$ as $\gamma\rightarrow0$. Then, for
each of the above cases 1-4, for sufficiently small $\gamma>0$, the smooth
flow has transverse homoclinic orbits which limit to the billiard's transverse
homoclinic orbits for all $0<\epsilon<\epsilon_{0}(\gamma)$.
\end{prop}

Indeed, for sufficiently small $\gamma>0$ equation
(\ref{perturbation of geometry}) is valid, and thus the homoclinic billiard
orbit $P^{b-\hom-\gamma}$ is transverse, so the above theorem follows
immediately from Theorem \ref{thm:persistance} and the discussion after it.
Based on this proposition and Propositions \ref{exponential}-\ref{coulomb} we conclude:

\begin{prop}
\label{prop:ellip-pot}Consider the Hamiltonian flow (\ref{eq:hambilellipse}),
where $V(q;\gamma,\epsilon)$ is a billiard potential limiting to the billiard
flow in $D_{\gamma}$ ($V(q;\gamma,\epsilon)$ satisfies conditions I-IV for all
$\gamma$ values). Further assume that the potential $V(q;\gamma,\epsilon)$ is
boundary dominated and is given near the boundary of $D_{\gamma}$ by
$W(Q;\epsilon),$ so that (\ref{assumption on m}) holds for the corresponding
$\delta$ values which are specified bellow. Then, for each of the above cases
1-4, for sufficiently small $\gamma>0$, the smooth flow has transverse
homoclinic orbits which limit to the billiard's transverse homoclinic orbits
and thus is non-integrable for all $0<\epsilon<\epsilon_{0}(\gamma)$, where

\begin{itemize}
\item For $W(Q;\epsilon)=e^{-\frac{Q}{\epsilon}}$ : $\delta=O(-\epsilon
\ln\epsilon)$ and $\epsilon_{0}(\gamma)=\gamma^{3+\varkappa},\varkappa>0.$

\item For $W(Q;\epsilon)=e^{-\frac{Q^{2}}{\epsilon}}$ : $\delta=O(\sqrt
{-\epsilon\ln\epsilon})$ and $\epsilon_{0}(\gamma)=\gamma^{6+\varkappa
},\varkappa>0.$

\item For $W(Q;\epsilon)=(\frac{\epsilon}{Q})^{\alpha}$ : $\delta
=O(\sqrt[3+\frac{1}{\alpha}]{\epsilon})$ and $\epsilon_{0}(\gamma
)=\gamma^{3+\frac{1}{\alpha}+\varkappa},\varkappa>0$.
\end{itemize}
\end{prop}

\section{Discussion}

The paper includes three main results:

\begin{itemize}
\item Theorems \ref{Main th0}-\ref{Main thr} deal with the smooth convergence
of flows in steep potentials to the billiard's flow in the multi-dimensional
case. These results, which are a natural extension of \cite{TuRK98}, provide a
powerful theoretical tool for proving the persistence of various billiard
trajectories in the smooth systems, and vice versa. The unavoidable emergence
of degenerate tangencies in the higher dimensional setting, and the study of
corners and regular tangencies (extending \cite{RKTu99},\cite{turk03} to
higher dimensions) have yet to be addressed.

\item Theorems \ref{thm-est}-\ref{thim} provide the first order corrections
for approximating the smooth flows by billiards for regular reflections.
Theorem \ref{thm-est} proposes the appropriate zeroth order billiard geometry
which best approximates the steep billiard and a simple formula for computing
the first order correction terms, thus allowing to study the effect of
smoothing. The smooth flow and the billiard flow do not match in a boundary
layer - the width of it and the time spent in it are specified in Theorem
\ref{thim}. Propositions \ref{exponential}-\ref{coulomb} supply the estimates
for the boundary layer width and the accuracy of the auxiliary billiard
approximation for some typical potentials (exponential, Gaussian and
power-law). All these results are novel for any dimension, and propose a new
approach for studying problems with relatively steep potentials. A plethora of
questions regarding the differences between the smooth and hard wall systems
can now be rigorously analyzed.

\item Theorem \ref{thm:persistance} and Proposition \ref{prop:ellip-pot}: The
above mentioned $C^{1}$ estimates of the error terms lead naturally to the
persistence Theorem \ref{thm:persistance}. Applying these results to the
billiards studied in \cite{DFRR01}, we prove that the motion in steep
potentials in various deformed ellipsoids are non-integrable for an open
interval of the steepness parameter, and we provide a lower bound for this
interval length for the above mentioned typical potentials. While the analysis
of higher dimensional Hamiltonian systems is highly non-trivial, we
demonstrate here that some results which are obtained for maps may be
immediately extended to the smooth steep case. We note that the same statement
works in the opposite direction. Furthermore, one may use the first order
corrections developed in Theorem \ref{thm-est} and Propositions
\ref{exponential}-\ref{coulomb} to study the possible appearance of
non-integrabilty due to the introduction of smooth potentials.
\end{itemize}

\bigskip

As mentioned before, these results may give the impression that the smooth
flow and the billiard flow are indeed very similar. While in this work we
emphasize the closeness of the two flows, it is important to bear in mind that
this is not the case in general. This observation applies to the local
behavior near solutions which are not structurally stable and is especially
important when dealing with asymptotic properties such as ergodicity, as
discussed below.

Let us first remark about the local behavior. First, as in the two-dimensional
settings, we expect that singular orbits or polygons of the billiard give rise
to various types of orbits in the smooth setting. The larger the dimension of
the system, the larger is the variety of orbits which may emerge from these
singularities. Moreover, in this higher dimensional setting, even though our
theory implies that regular elliptic or partially-elliptic periodic orbits
persist, the motion near them (and their stability) may change.

Global properties of the phase space are even more sensitive to small changes.
If the billiard periodic orbit is hyperbolic, while it and its local stable
and unstable manifolds persist (see for example Theorem \ref{thm:persistance}%
), their global structure in the smooth case may be quite different; First of
all, integrability of one of the systems does not imply integrability of the
other (for example, it may be possible to use the correction terms computed in
Section \ref{sec:crestimates} to establish that the smooth flow has separatrix
splitting even when the billiard is integrable). Second, if the billiard flow
has singularities, the global manifolds of a hyperbolic billiard orbit may
have discontinuities and singularities whereas the global manifolds of the
smooth orbit are smooth (see for example \cite{TuRK98}).

Finally, the most celebrated global property one is interested in is
ergodicity and mixing. Indeed, Boltzmann suggested that the gas molecules
interacting in a box should have a fast decaying correlation function and
proposed the analogy of the corresponding dispersing hard balls system. In
modern terminology, Boltzmann claimed that for \emph{sufficiently large}
systems the hard sphere gases are ergodic and mixing \cite{Kry79} and hence so
are the real gases. Sinai \cite{Sina63} proved that the dispersion property is
sufficient for proving that the system of two disks on a two-torus is ergodic
and mixing, and following this fundamental work the study of the dynamics and
mixing properties in various two-dimensional billiard tables had flourished
\cite{Bu79,Bu01,Wo86,GarGal94} (the behavior of billiards in higher dimensions
is much less studied, see \cite{Wo90,BuRe98f,BuRe98,PrSm00}\cite{SiCh87}%
\cite{SiNa04} and references therein).

The suspicion that the motion in smooth steep potentials may have a different
character has been lurking all along. In fact, several works where dedicated
to proving that in some cases (finite-range axis-symmetric potentials) the
motion may be still ergodic \cite{Sina63,Ku76,KuMu81,DoLi91,BaTo04}. In
\cite{do99} it was shown that when two particles with a finite range potential
move on a two-dimensional torus stable periodic orbit may emerge. In
\cite{TuRK98} we proved that in the two dimensional case ($C^{r}$ smooth
potentials, not necessarily finite range, not necessarily symmetric), near
singular trajectories (tangent trajectories or corner trajectories) new
islands of stability are born in the smooth flow for \emph{arbitrarily steep}
potentials. Thus there is a fundamental difference in the ergodic properties
of hard-wall potentials as compared to smooth potentials. Although these
results only apply to two-particle systems, they raise the possibility that
systems with large numbers of particles interacting by smooth potentials could
also be non-ergodic. The tools developed here may be useful in studying these
possibilities.\bigskip

\section{Acknowledgments}
This research is supported by the Israel Science Foundation (Grant
no. 926/04) and by the Minerva foundation.

\section{Appendix}

\subsection{Proof of Theorems \ref{Main th0} and \ref{Main thr}}

\label{main} By Condition I the Hamiltonian flow is $C^{r}$-close to the
billiard flow outside an arbitrarily small boundary layer. So we will
concentrate our attention on the behavior of the Hamiltonian flow inside such
a layer.

\begin{figure}[ptb]
\centering \psfig{figure =
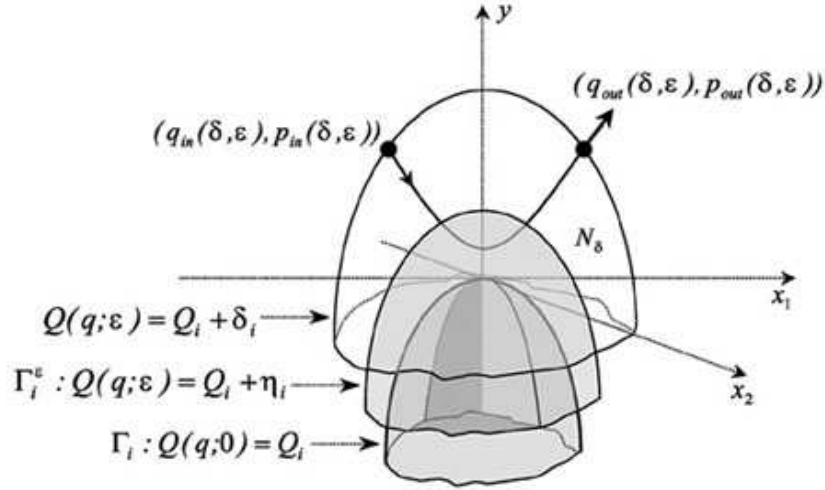,height=70mm,width=110mm}\caption{Hamiltonian flow inside small
boundary layer.}%
\label{fig:hamiltonian}%
\end{figure}

Let the initial conditions correspond to the billiard orbit which hits a
boundary surface $\Gamma_{i}$ at a (non-corner) point $q_{c}$. By Condition
IIa, the surface $\Gamma_{i}$ is given by the equation $Q(q;0)=Q_{i}$, hence
the boundary layer near $\Gamma_{i}$ can be defined as $N_{\delta
}=\{|Q(q;\epsilon)-Q_{i}|\leq\delta\}$, where $\delta$ tends to zero
sufficiently slowly as $\epsilon\rightarrow+0$. Take $\epsilon$ sufficiently
small. The smooth trajectory enters $N_{\delta}$ at some time $t_{in}%
(\delta,\epsilon)$ at a point $q_{in}(\delta,\epsilon)$ which is close to the
collision point $q_{c}$ with the velocity $p_{in}(\delta,\epsilon)$ which is
close to the initial velocity $p_{0}$. See Figure \ref{fig:hamiltonian}. The
same trajectory exits from $N_{\delta}$ at the time $t_{out}(\delta,\epsilon)$
at a point $q_{out}(\delta,\epsilon)$ with velocity $p_{out}(\delta,\epsilon
)$. In these settings, the theorems are equivalent ($r=0$ corresponds to
Theorem \ref{Main th0}, while $r>0$ corresponds to Theorem \ref{Main thr}) to
proving the following statements:%
\begin{equation}
\lim_{\delta\rightarrow0}\lim_{\epsilon\rightarrow+0}\bigg\|\bigg(q_{out}%
(\delta,\epsilon),t_{out}(\delta,\epsilon)\bigg)-\bigg(q_{in}(\delta
,\epsilon),t_{in}(\delta,\epsilon)\bigg)\bigg\|_{C^{r}}=0, \label{Travel is 0}%
\end{equation}
which guarantees that the trajectory does not travel along the boundary, and
(see (\ref{Reflection}))
\begin{equation}
\lim_{\delta\rightarrow0}\lim_{\epsilon\rightarrow+0}\bigg\|p_{out}%
(\delta,\epsilon)-p_{in}(\delta,\epsilon)+2n(q_{in})\langle p_{in}%
(\delta,\epsilon),n(q_{in})\rangle\bigg\|_{C^{r}}=0, \label{Correct normal}%
\end{equation}
where $n(q)$ is the unit inward normal to the level surface of $Q$ at the
point $q$.

With no loss of generality, assume that $Q(q;0)$ increases as $q$ leaves
$D^{\prime}s$ boundary towards $D^{\prime}s$ interior. Choose the coordinates
$(x,y)$ so that the hyperplane $x$ is tangent to the level surface
$Q(q;\epsilon)=Q(q_{c};\epsilon)$ and the $y$-axis is the inward normal to
this surface at $q=q_{c}$. Hence, the partial derivatives of $Q$ satisfy:%
\begin{equation}
Q_{x}|_{(q_{c};\epsilon)}=0,\;\;\;\;Q_{y}|_{(q_{c};\epsilon)}=1
\label{Partial der of Q}%
\end{equation}
By (\ref{Hamiltonian epsilon}) and Condition II, near the boundary the
equations of motion have the form:
\begin{equation}
\dot{x}=\frac{\partial H}{\partial p_{x}}=p_{x}\;\;\;\;\;\;\;\dot{p_{x}%
}=-\frac{\partial H}{\partial x}=-W^{\prime}(Q;\epsilon)Q_{x},
\label{Equations of motion x}%
\end{equation}%
\begin{equation}
\dot{y}=\frac{\partial H}{\partial p_{y}}=p_{y}\;\;\;\;\;\;\;\dot{p_{y}%
}=-\frac{\partial H}{\partial y}=-W^{\prime}(Q;\epsilon)Q_{y}.
\label{Equations of motion y}%
\end{equation}
We start with the $C^{0}$ version of (\ref{Travel is 0}) and
(\ref{Correct normal}). First, we will prove that given a sufficiently slowly
tending to zero $\xi(\epsilon)$, if the orbit stays in the boundary layer
$N_{\delta}$ for all $t\in\lbrack t_{in},t_{in}+\xi]$, then in this time
interval%
\begin{equation}
q(t)=q_{in}(\delta,\epsilon)+O(\xi), \label{Approx for q}%
\end{equation}%
\begin{equation}
p_{x}(t)=p_{x}(t_{in}(\delta,\epsilon))+O(\xi), \label{Approx for p}%
\end{equation}%
\begin{equation}
\frac{p_{y}(t)^{2}}{2}+W(Q(q(t);\epsilon);\epsilon)=\frac{p_{y}(t_{in}%
(\delta,\epsilon))^{2}}{2}+W(\delta;\epsilon)+O(\xi).
\label{Approx for conserv}%
\end{equation}
Note that (\ref{Approx for q}) follows immediately from
(\ref{Equations of motion x})-(\ref{Equations of motion y}) and the fact that
$p$ is uniformly bounded by the energy constraint $\frac{p^{2}}{2}%
=H-W(Q;\epsilon)\leq H=\frac{1}{2}$. In fact, $q_{in}-q_{c}$ tends to zero as
$O(\delta)$ for regular trajectories and $O(\sqrt{\delta})$ for non-degenerate
tangent trajectories, so by assuming that $\xi(\epsilon)$ is slow enough, we
extract from (\ref{Approx for q}) that
\begin{equation}
q(t)=q_{c}+O(\xi). \label{afq}%
\end{equation}
Now, from (\ref{Partial der of Q}), (\ref{afq}), for $t\in\lbrack
t_{in}(\delta,\epsilon),t_{in}(\delta,\epsilon)+\xi]$ we have
\begin{equation}
Q_{x}(q(t);\epsilon)=O(\xi),\;\;Q_{y}(q(t);\epsilon)=1+O(\xi).
\label{Q - partial der}%
\end{equation}
Divide the interval $I=[t_{in},t_{in}+\xi]$ into two sets: $I_{<}$ where
$|W^{\prime}(Q;\epsilon)|<1$ and $I_{>}$ where $|W^{\prime}(Q;\epsilon)|\geq
1$. In $I_{<}$ we have $\dot{p_{x}}=O(\xi)$ by (\ref{Equations of motion x}%
),(\ref{Q - partial der}). In $I_{>}$, as $|W^{\prime}(Q;\epsilon)|\geq1$ and
$Q_{y}\neq0$, we have that $\dot{p_{y}}$ is bounded away from zero, so in
(\ref{Equations of motion x}) we can divide $\dot{p_{x}}$ by $\dot{p_{y}}$:
\[
\frac{dp_{x}}{dp_{y}}=\frac{Q_{x}}{Q_{y}}.
\]
It follows that the change in $p_{x}$ on $I$ can be estimated from above as
$O(\xi^{2})$ (the contribution from $I_{<}$) plus $O(\xi)$ times the total
variation in $p_{y}$. Thus, in order to prove (\ref{Approx for p}), it is
enough to show that the the total variation in $p_{y}$ on $I$ is uniformly
bounded. Recall that $p_{y}$ is uniformly bounded ($|p_{y}|\leq1$ from the
energy constraint) and monotone (as $W^{\prime}(Q)<0$ and $Q_{y}>0$, we have
$\dot{p}_{y}>0$, see (\ref{Equations of motion y})) everywhere on $I$, so its
total variation is uniformly bounded indeed. Thus, (\ref{Approx for p}) is
proven. The approximate conservation law (\ref{Approx for conserv}) follows
now from (\ref{Approx for p}) and the conservation of $H=\frac{p_{y}^{2}}%
{2}+\frac{p_{x}^{2}}{2}+W(Q(q;\epsilon);\epsilon)$.

Finally, we prove that $\tau_{\delta}$, the time the trajectory spends in the
boundary layer $N_{\delta}$, tends to zero as $\epsilon\rightarrow0$. This
step completes the proof of Theorem \ref{Main th0}: by plugging the time
$\tau_{\delta}\rightarrow0$ instead of $\xi$ in the right-hand sides of
(\ref{Approx for q}),(\ref{Approx for p}),(\ref{Approx for conserv}), we
immediately obtain the $C^{0}$-version of (\ref{Travel is 0}) and
(\ref{Correct normal}).

Let us start with the non-tangent case, i.e. with the trajectories such that
$p_{y}(t_{in})$ is bounded away from zero. From Condition III it follows that
the value of $W_{in}=W_{out}=W(Q=\delta;\epsilon)$ vanishes as $\epsilon
\rightarrow+0$. Hence, by (\ref{Approx for conserv}) the momentum $p_{y}(t)$
stays bounded away from zero as long as the potential $W(Q;\epsilon)$ remains
small. Choose some small $\nu$, and divide $N_{\delta}$ into two parts
$N_{<}:=\{W:W(Q;\epsilon)\leq\nu\}$ and $N_{>}=\{W:W(Q;\epsilon)>\nu\}$.
First, the trajectory enters $N_{<}$. Since the value of $\frac{d}%
{dt}Q(q)=p_{x}Q_{x}+p_{y}Q_{y}$ is negative and bounded away from zero in
$N_{<}$ (because $Q_{x}$ is small, and $p_{y}$ and $Q_{y}$ are non-zero), the
trajectory must reach the inner part $N_{>}$ by a time proportional to the
width of $N_{<}$, which is $O(\delta)$. Also, we can conclude that if the
trajectory leaves $N_{>}$ after some time $t_{>}$, it must have $p_{y}>0$ and,
arguing as above, we obtain that $t_{out}-t_{in}=O(\delta)+t_{>}$. Let us show
that $t_{>}\rightarrow0$ as $\epsilon\rightarrow+0$. Using
(\ref{Equations of motion y}), the fact that the total variation of $p_{y}$ is
bounded, and Condition IV, we obtain
\[
|t_{>}|\leq\frac{C}{\min_{N_{>}}|W^{\prime}(Q;\epsilon)|}=C\max_{N_{>}%
}|{\mathcal{Q}}^{\prime}(W;\epsilon)|\rightarrow0\;\;\;\mathrm{as}%
\;\;\epsilon\rightarrow+0.
\]
So, in the non-tangent case, the collision time is $O(\delta+t_{>})$, i.e. it
tends to zero indeed.

This result holds true for $p_{y,in}$ bounded away from zero, and it remains
valid for $p_{y,in}$ tending to zero sufficiently slowly. Hence, we are left
with the case where $p_{y,in}$ tends to zero as $\epsilon\rightarrow0$ (the
case of nearly tangent trajectories). Inside $N_{\delta}$, since $W$ is
monotone by (\ref{Derivative of W}), we have $W(Q;\epsilon)>W_{in}%
=W(\delta;\epsilon)$. Therefore, by (\ref{Approx for conserv}), $p_{y}(t)$
stays small unless the trajectory leaves $N_{\delta}$ or $t-t_{in}$ becomes
larger than a certain bounded away from zero value. From (\ref{Approx for p})
it follows then that $p_{x}(t)$ remains bounded away from zero. By
(\ref{Equations of motion x}),(\ref{Equations of motion y}),
\[
\dot{Q}:=\frac{d}{dt}Q(q(t);\epsilon)=Q_{x}p_{x}+Q_{y}p_{y}%
\]
so $\dot{Q}$ is small, yet%

\[
\frac{d^{2}}{dt^{2}}Q(q(t);\epsilon)=p_{x}^{T} Q_{xx}p_{x}+2Q_{xy}p_{x}%
p_{y}+Q_{yy}p_{y}^{2}-W^{\prime}(Q;\epsilon)(Q_{x}^{2}+Q_{y}^{2}).
\]
For a non-degenerate tangency, $p_{x}^{T} Q_{xx}p_{x}$ is positive and bounded
away from zero. Therefore, as $p_{y}$ is small and $W^{\prime}(Q;\epsilon)$ is
negative, we obtain that $\frac{d^{2}}{dt^{2}}Q(q(t);\epsilon)$ is positive
and bounded away from zero for a bounded away from zero interval of time
(starting with $t_{in}$). It follows that
\begin{equation}
Q(q(t);\epsilon)\geq Q(q_{in};\epsilon)+\dot{Q}(t_{in})(t-t_{in}%
)+C(t-t_{in})^{2} \label{qdd}%
\end{equation}
on this interval, for some constant $C>0$. We see from (\ref{qdd}), that the
trajectory has to leave the boundary layer $N_{\delta}=\{|Q(q;\epsilon
)-Q_{i}|\leq\delta=|Q(q_{in};\epsilon)-Q_{i}|\}$ in a time of order $O(\dot
{Q}(t_{in}))=O(Q_{x}(q_{in}))+O(p_{y,in})=O(q_{in}-q_{c})+O(p_{y,in})$. As
$q_{in}-q_{c}=O(\sqrt{\delta})$ for a non-degenerate tangency, we see that the
time the nearly-tangent orbit may spend in the boundary layer is
$O(\sqrt{\delta}+p_{y,in})$, i.e. in this case it tends to zero as well. This
completes the proof of Theorem \ref{Main th0}.

Now we prove Theorem \ref{Main thr} - the $C^{r}$-convergence for the
non-tangent case. Again, divide $N_{\delta}$ into $N_{<}$ and $N_{>}$ for a
small $\nu$ and consider the limit $\lim_{\delta\rightarrow0}\lim
_{\nu\rightarrow0}\lim_{\epsilon\rightarrow+0}$. As we have shown above,
$\dot{Q}\neq0$ in $N_{<}$, thus we can divide the equations of motion
(\ref{Equations of motion x}), (\ref{Equations of motion y}) by $\dot{Q}$:
\begin{align}
\frac{dq}{dQ}  &  =\frac{p}{Q_{x}p_{x}+p_{y}Q_{y}}\nonumber\\
\frac{dp}{dQ}  &  =-W^{\prime}(Q;\epsilon)\frac{\nabla Q}{Q_{x}p_{x}%
+p_{y}Q_{y}},\label{Divided equqtion of motion}\\
\frac{dt}{dQ}  &  =\frac{1}{Q_{x}p_{x}+p_{y}Q_{y}}\nonumber
\end{align}
Equations (\ref{Divided equqtion of motion}) can be rewritten in an integral
form:%
\[
q(Q_{2})-q(Q_{1})=\int_{Q_{1}}^{Q_{2}}F_{q}(q,p)dQ,
\]%
\begin{equation}
p(Q_{2})-p(Q_{1})=-\int_{W(Q_{1})}^{W(Q_{2})}F_{p}(q,p)dW(Q),
\label{Integral form}%
\end{equation}%
\[
t(Q_{2})-t(Q_{1})=\int_{Q_{1}}^{Q_{2}}F_{t}(q,p)dQ,
\]
where $F_{q},F_{p}$ and $F_{t}$ denote some functions of $(q,p)$ which are
uniformly bounded along with all derivatives. In $N_{<}$, the change in $Q$ is
bounded by $\delta$ and the change in $W$ is bounded by $\nu$. Hence, the
integrals on the right-hand side are small. Applying the successive
approximation method, we obtain that the Poincar\'{e} map (the solution to
(\ref{Integral form})) from $Q=Q_{1}$ to $Q=Q_{2}$ limits to the identity map
(along with all derivatives with respect to initial conditions) as $\delta
,\nu\rightarrow0$. It follows that in order to prove the theorem, i.e. to
prove (\ref{Travel is 0}),(\ref{Correct normal}), we need to prove
\begin{equation}
\lim_{\nu\rightarrow0}\lim_{\epsilon\rightarrow+0}\bigg\|\bigg(q_{out}%
,t_{out}\bigg)-\bigg(q_{in},t_{in})\bigg)\bigg\|_{C^{r}}=0,
\label{Travel is nu}%
\end{equation}
and
\begin{equation}
\lim_{\nu\rightarrow0}\lim_{\epsilon\rightarrow+0}\bigg\|p_{out}%
-p_{in}+2n(q_{in})\langle p_{in},n(q_{in})\rangle\bigg\|_{C^{r}}=0,
\label{Correct nu}%
\end{equation}
where $(q_{in},p_{in},t_{in})$ and $(q_{out},p_{out},t_{out})$ correspond now
to the intersections of the orbit with the cross-section $W(Q(q,\epsilon
),\epsilon)=\nu$. By Condition IV, as $\epsilon\rightarrow0$ the function
${\mathcal{Q}}(W;\epsilon)$ tends to zero uniformly along with all its
derivatives in the region $\nu\leq W\leq H$ for any $\nu$ bounded away from
zero. Therefore, the same holds true for a sufficiently slowly tending to zero
$\nu$ and $W^{\prime}(Q;\epsilon)=({\mathcal{Q}}^{\prime}(W;\epsilon))^{-1}$
is bounded away from zero in the region $N_{>}$. Hence, by
(\ref{Equations of motion y}), the derivative $\dot{p_{y}}$ is bounded away
from zero as well. Therefore, we can divide the equations of motion
(\ref{Equations of motion x}),(\ref{Equations of motion x}) by $\frac{dp_{y}%
}{dt}$:%
\begin{equation}
\frac{dq}{dp_{y}}=-\mathcal{Q}^{\prime}(W;\epsilon)\frac{p}{Q_{y}%
},\;\;\;\;\;\frac{dt}{dp_{y}}=-\mathcal{Q}^{\prime}(W;\epsilon)\frac{1}{Q_{y}%
},\;\;\;\;\;\frac{dp_{x}}{dp_{y}}=\frac{Q_{x}}{Q_{y}}, \label{dx-dz equation}%
\end{equation}
where
\begin{equation}
W=H-\frac{1}{2}p^{2}. \label{W from conserv}%
\end{equation}
Condition IV implies that the $C^{r}$-limit as $\epsilon\rightarrow0$ of
(\ref{dx-dz equation}) is%
\begin{equation}
\frac{d(q,t)}{dp_{y}}=0,\;\;\;\;\;\frac{dp_{x}}{dp_{y}}=\frac{Q_{x}}{Q_{y}}
\label{Cr limit}%
\end{equation}
Since the change in $p_{y}$ is finite and the functions on the right-hand side
of (\ref{dx-dz equation}) are all bounded, the solution of this system is the
$C^{r}$-limit of the solution of (\ref{dx-dz equation}). From (\ref{Cr limit})
we obtain that in the limit $\epsilon\rightarrow0$ $(q_{in},t_{in}%
)=(q_{out},t_{out})$, so (\ref{Travel is nu}) is proved. Second, we obtain
from (\ref{Cr limit}) that%
\[
(p_{x,out}-p_{x,in})Q_{y}(q_{in};\epsilon)=(p_{y,out}-p_{y,in})Q_{x}%
(q_{in};\epsilon)
\]
in the limit $\epsilon\rightarrow0$, which, in the coordinate independent
vector notation (see e.g. \ref{eq:normal-py}), and by using $(q_{in}%
,t_{in})=(q_{out},t_{out}),$ amounts to the correct reflection law.

\subsection{\label{sec:Picard}\textbf{Picard iteration for equations with
small right-hand side.}}

Before we proceed to the proof of Lemmas \ref{freeflight} and
\ref{error_of_reflection}, we recall the main tool of their proofs - the
Picard iteration scheme for equations with small right hand side. Consider the
differential equation
\begin{equation}
\dot{v}=\psi(v,\mu,t,\epsilon) \label{psiv}%
\end{equation}
where $\psi$ is a $C^{r}$-smooth function of $v$ and $\mu$, continuous with
respect to $t$ and $\epsilon$. Assume that for $t\in\lbrack0,L(\epsilon)]$ and
bounded $(v,\mu)$ we have a function $J(\epsilon)$ such that $J(\epsilon
)L(\epsilon)\rightarrow0$ and
\begin{equation}
\Vert\psi\Vert_{C^{r}}\leq J(\epsilon). \label{equ}%
\end{equation}
Then, according to the contraction mapping principle, the Picard iterations
$v_{n}$ where
\begin{equation}
v_{n+1}(t)=v_{0}+\int_{0}^{t}\psi(v_{n}(s),\mu,s,\epsilon)ds \label{pic}%
\end{equation}
converge to the solution of (\ref{psiv}) starting at $t=0$ with initial
condition $v(0)=v_{0}$ on the interval $t\in\lbrack0,L(\epsilon)]$, in the
$C^{r}$-norm as a function of $v_{0}$ and $\mu$:
\[
v_{n}(t;v_{0},\mu)\rightarrow_{_{C^{r}}}v(t;v_{0},\mu)=v_{0}+\int_{0}^{t}%
\psi(v(s;v_{0},\mu),\mu,s,\epsilon)ds.
\]
One can show by induction that $\Vert v_{n}(t)-v_{0}\Vert_{_{C^{r}}%
}=O(L(\epsilon)J(\epsilon))$ uniformly for all $n$. Then it follows that
\begin{equation}
v(t;v_{0},\mu)=v_{0}+O_{_{C^{r}}}(L(\epsilon)J(\epsilon)). \label{pic3}%
\end{equation}

It is easy to show that
\begin{equation}
v(t;v_{0},\mu)=v_{n}(t;v_{0},\mu)+O_{_{C^{r-1}}}((L(\epsilon)J(\epsilon
))^{n+1}) \label{est}%
\end{equation}
(such kind of estimates are, in fact, a standard tool in the averaging
theory). In order to prove (\ref{est}), we will use induction in $n$. At $n=0$
we have even better result than (\ref{est}) (see (\ref{pic3})). Now note that
\begin{align*}
v(t)-v_{n+1}(t)  &  =\int_{0}^{t}(\psi(v(s),\mu,s,\epsilon)-\psi(v_{n}%
(s),\mu,s,\epsilon)ds\\
&  =\int_{0}^{t}\left(  \int_{0}^{1}\psi_{v}^{\prime}(v_{n}(s)+z(v(s)-v_{n}%
(s)),\mu,s,\epsilon)dz\right)  \cdot(v(s)-v_{n}(s))ds.
\end{align*}
It follows immediately that
\[
\Vert v-v_{n+1}\Vert_{_{C^{r-1}}}=O(L(\epsilon)\Vert\psi_{v}^{\prime}%
\Vert_{_{C^{r-1}}})\cdot O(\Vert v-v_{n}\Vert_{_{C^{r-1}}})=O(L(\epsilon
)J(\epsilon))\cdot\Vert v-v_{n}\Vert_{_{C^{r-1}}},
\]
and (\ref{est}) indeed holds true by induction.

\subsection{\label{sec:proof of frreflights}Proof of Lemma \ref{freeflight}}

The \textquotedblleft free flight\textquotedblright\ (the motion inside
$D^{\epsilon}$) is composed of motion in $D_{int}^{\epsilon}$ (the region
outside of $N_{\delta}$) and the motion in the layer $N_{<}=D^{\epsilon
}\backslash D_{int}^{\epsilon}$. We show that in each of these regions the
equations may be brought to the form (\ref{psiv}),(\ref{equ}). We will first
consider the flight inside $D_{int}^{\epsilon}$. Recall that the equations of
motion for the smooth orbit are
\begin{equation}%
\begin{array}
[c]{l}%
\dot{q}=p\\
\dot{p}=-\nabla V(q;\epsilon).
\end{array}
\label{gradV}%
\end{equation}
Let us make the following change of coordinates
\begin{equation}
\tilde{q}(t):=q(t)-p(t)t \label{change of coordinates}%
\end{equation}
Then (\ref{gradV}) takes the form%
\begin{equation}%
\begin{array}
[c]{l}%
\dot{\tilde{q}}=\nabla V(\tilde{q}+pt;\epsilon)t\\
\dot{p}=-\nabla V(\tilde{q}+pt;\epsilon)
\end{array}
\label{out}%
\end{equation}
with initial data $(\tilde{q}(0),p(0))=(q_{0},p_{0})$. Since the time spent in
$D_{int}^{\epsilon}$ must be finite as it is $C^{r}-$close to the billiard's
travel time in $D_{int}^{\epsilon}$ which is finite here, and using
(\ref{m-small}), we have%
\[
\Vert\psi\Vert_{C^{r}}=\Vert%
\begin{pmatrix}
\nabla V(\tilde{q}+pt;\epsilon)t\\
-\nabla V(\tilde{q}+pt;\epsilon)
\end{pmatrix}
\Vert_{C^{r}}=O(m^{(r)}(\delta(\epsilon);\epsilon)).
\]
Thus, system (\ref{out}) does satisfy (\ref{equ}) with $L=O(1)$,
$J=O(m^{(r)})$. It follows then from (\ref{pic3}) that
\begin{equation}
p(t)=p_{0}+O_{_{C^{r}}}(m^{(r)}). \label{pff}%
\end{equation}
Furthermore, by applying $n=1$ Picard iteration (\ref{pic}), we obtain from
(\ref{est}) the following estimate for $p(t)$:
\begin{equation}
p(t)=p_{0}-\int_{0}^{t}\nabla V(q_{0}+p_{0}s;\epsilon)ds+O_{_{C^{r-1}}%
}((m^{(r)})^{2}). \label{pf}%
\end{equation}
By integrating the equation $\dot{q}=p$, we also obtain from (\ref{pff}) that
\begin{equation}
q(t)=q_{0}+p_{0}t+O_{_{C^{r}}}(m^{(r)}). \label{qff}%
\end{equation}
Next, we show that the equations in the layer $N_{<}=\{W:W(Q;\epsilon)\leq
\nu\}$ can be brought to the form (\ref{psiv}),(\ref{equ}) as well. Recall
(see the proof of Theorem \ref{Main thr}) that $\dot{Q}=\langle\nabla
Q,p\rangle$ is bounded away from zero in $N_{<}$, hence $Q$ can be taken as a
new independent variable (it changes in the interval $\eta\leq Q-Q_{i}%
\leq\delta$). Now the time $t$ is considered as a function of $Q$ and of the
initial conditions $(q(t_{\delta}),p(t_{\delta})$ (where $t_{\delta}$ is the
moment the trajectory enters $N_{<}$). Recall that we showed in the proof of
Theorem \ref{Main thr} that $t$ is a smooth function of the initial
conditions, with all the derivatives bounded. So, in $N_{<}$, we rewrite
(\ref{out}) as
\[%
\begin{array}
[c]{l}%
\frac{d\tilde{q}}{dQ}=W^{\prime}(Q;\epsilon)\frac{\nabla Q(\tilde
{q}+pt;\epsilon)}{\langle\nabla Q(\tilde{q}+pt;\epsilon),p\rangle}t\\
\frac{dp}{dQ}=-W^{\prime}(Q;\epsilon)\frac{\nabla Q(\tilde{q}+pt;\epsilon
)}{\langle\nabla Q(\tilde{q}+pt;\epsilon),p\rangle}.
\end{array}
\]
As $W$ is a monotone function of $Q$ (i.e. $W^{\prime}(Q;\epsilon)\neq0$), we
can take $W$ as a new independent variable, so the equations of motion will
take the form
\begin{equation}%
\begin{array}
[c]{l}%
\frac{d\tilde{q}}{dW}=\frac{\nabla Q(\tilde{q}+pt;\epsilon)}{\langle\nabla
Q(\tilde{q}+pt;\epsilon),p\rangle}t\\
\frac{dp}{dW}=-\frac{\nabla Q(\tilde{q}+pt;\epsilon)}{\langle\nabla
Q(\tilde{q}+pt;\epsilon),p\rangle}.
\end{array}
\label{eqw}%
\end{equation}
Since all the derivatives of $t$ with respect to the initial conditions are
bounded, we may consider (\ref{eqw}) as the system of type (\ref{psiv}%
),(\ref{equ}) with $J=O(1)$, and $L=O(\nu)$ (recall that the value of $W$
changes monotonically from $W_{0}=W(\delta,\epsilon)$ to $\nu$). Thus, by
applying one Picard iteration (\ref{pic}), we obtain from (\ref{est}) that
\[
p(W)=p(W_{0})-\int_{W(\delta,\epsilon)}^{W}\frac{\nabla Q(\tilde{q(W_{0}%
)}+p(W_{0})t;\epsilon)}{\langle\nabla Q(\tilde{q}(W_{0})+p(W_{0}%
)t;\epsilon),p(W_{0})\rangle}dW+O_{_{C^{r-1}}}(\nu^{2}).
\]
From (\ref{pic3}) we also obtain
\[
p(W)=p(W_{0})+O_{_{C^{r}}}(\nu).
\]
Note that $O_{_{C^{r-1}}}(\nu^{2})$ and $O_{_{C^{r}}}(\nu)$ refer here to the
derivatives (with respect to the initial conditions) of $p$ at constant $W$
or, equivalently, at constant $Q$. Returning to the original time variable,
these equations yield
\[
p(t)=p(t_{\delta})+O_{_{C^{r}}}(\nu)=p(t_{\delta})-\int_{t_{\delta}}^{t}\nabla
V(q(t_{\delta})+p(t_{\delta})(s-t_{\delta});\epsilon)ds+O_{C^{r-1}}(\nu^{2}).
\]
Using expressions (\ref{pff}),(\ref{pf}) for $p(t_{\delta})$ and (\ref{qff})
and $q(t_{\delta})$, we finally obtain
\begin{equation}
p(t)=p_{0}+O_{_{C^{r}}}(\nu+m^{(r)})=p_{0}-\int_{0}^{t}\nabla V(q_{0}%
+p_{0}s;\epsilon)ds+O_{C^{r-1}}((m^{(r)}+\nu)^{2}) \label{pf0}%
\end{equation}
for all $t$ such that $q(t)\in D^{\epsilon}$, in complete agreement with the
claim of the lemma (as we mentioned, the $O_{_{C^{r-1}}}(\cdot)$ and
$O_{_{C^{r}}}(\cdot)$ terms refer to the derivatives at constant $Q$). The
corresponding expression for $q(t)$ (see (\ref{lemfre})) is obtained by
integrating the equation $\dot{q}=p$. The expression (\ref{tauint}) for the
flight time $\tau$ is immediately found from the relation $W(Q(q(\tau
);\epsilon);\epsilon)=\nu$ or, equivalently, $Q(q(t);\epsilon)=Q_{i}+\eta$
(recall that $\dot{Q}$ is bounded away from zero in the layer $N_{<}$).

\subsection{Proof of Lemma \ref{error_of_reflection}}

Here we compute the reflection map $R^{\epsilon}:(q_{in},p_{in}):\mapsto
(q_{out},p_{out})$ defined by the smooth trajectories within the most inner
layer $N_{>}:\{W\geq\nu\}$. We put the origin of the coordinate system at the
point $q_{in}$ (corresponding to $q$ at Figure \ref{fig:reflection}) and
rotate the axes with $\epsilon$ so that the $y$-axis will coincide with the
inward normal to the surface $Q(q;\epsilon)=Q(q_{in};\epsilon)$ at the point
$q_{in}$ (corresponds to $n(q)$ at Figure \ref{fig:reflection}), the
$x$-coordinates will correspond to the tangent directions. It is easy to see
that in the notations of Lemma \ref{error_of_reflection} we have (the explicit
dependence on $\epsilon$ is suppressed for brevity)
\begin{equation}
K(q_{in})=Q_{xx}(q_{in})/Q_{y}(q_{in}),\qquad l(q_{in})=Q_{xy}(q_{in}%
)/Q_{y}(q_{in}). \label{klq}%
\end{equation}
As we have shown in the proof of Theorem \ref{Main thr}, $\frac{dp_{y}}{dt}$
is bounded away from zero in $N_{>}$; Hence, we may use $p_{y}$ as the new
independent variable (see (\ref{dx-dz equation})). In order to bring the
equations of motion to the required form with the small right hand side, we
make the additional transformation%
\begin{equation}
p_{x}\rightarrow\tilde{p}=p_{x}-\frac{Q_{x}(q)}{Q_{y}(q)}p_{y}.
\label{eq:ptilde}%
\end{equation}
Note that $Q_{x}(q_{in};\epsilon)=0$, hence (see (\ref{klq}))
\begin{equation}
\tilde{p}=p_{x}-K(q_{in})(x-x_{in})p_{y}-l(q_{in})(y-y_{in})p_{y}%
+O((q-q_{in})^{2}). \label{zamp0}%
\end{equation}
In particular
\begin{equation}
\tilde{p}(t_{in})=p_{x,in}. \label{zamp}%
\end{equation}
After the transformation, equations (\ref{dx-dz equation}) take the form
\begin{align}
\frac{dq}{dp_{y}}  &  =-\mathcal{Q}^{\prime}(\frac{1}{2}-\frac{1}{2}%
p^{2})\frac{p}{Q_{y}},\;\;\;\;\label{dx-dz py}\\
\;\frac{dt}{dp_{y}}  &  =-\mathcal{Q}^{\prime}(\frac{1}{2}-\frac{1}{2}%
p^{2})\frac{1}{Q_{y}},\;\\
\;\;\;\;\frac{d\tilde{p}}{dp_{y}}  &  =\mathcal{Q}^{\prime}(\frac{1}{2}%
-\frac{1}{2}p^{2})\frac{d}{dq}\left(  \frac{Q_{x}}{Q_{y}}\right)  \frac
{p}{Q_{y}}p_{y}.
\end{align}
Since $\mathcal{Q}^{\prime}(W;\epsilon)$ is small in the inner layer, these
equations belong to the class (\ref{psiv}),(\ref{equ}), with $J=O(M^{(r)})$
(see (\ref{M-big})) and $L=O(1)$ (the change in $p_{y}$ is bounded by the
energy constraint). Thus, by (\ref{pic3}), we obtain (see (\ref{zamp}))
\begin{equation}
(q,t,\tilde{p})=(q_{in},t_{in},p_{x,in})+O_{_{C^{r}}}(M^{(r)}). \label{nul}%
\end{equation}
Recall that $W(q_{out})=W(q_{in})$. Therefore, by energy conservation,
\begin{equation}
p_{x,in}^{2}+p_{y,in}^{2}=p_{x,out}^{2}+p_{y,out}^{2}, \label{ener}%
\end{equation}
so (\ref{nul}) implies
\begin{equation}
p_{y,out}=-p_{y,in}+O_{_{C^{r}}}(M^{(r)}). \label{nuly}%
\end{equation}
By (\ref{nul}), and by using $Q_{x}(q_{in},\epsilon)=0$, equations
(\ref{dx-dz py}) may be written up to $O_{_{C^{r-1}}}((M^{(r)})^{2})$-terms
as
\begin{align}
\frac{dq}{dp_{y}}  &  =-\mathcal{Q}^{\prime}(\frac{1}{2}(1-p_{x,in}^{2}%
-p_{y}^{2}))\frac{(p_{x,in},p_{y})}{Q_{y}(q_{in})},\;\;\label{dx-dz cut}\\
\;\;\;\frac{dt}{dp_{y}}  &  =-\mathcal{Q}^{\prime}(\frac{1}{2}(1-p_{x,in}%
^{2}-p_{y}^{2}))\frac{1}{Q_{y}(q_{in})},\;\;\;\\
\;\;\frac{d\tilde{p}}{dp_{y}}  &  =\mathcal{Q}^{\prime}(\frac{1}{2}%
(1-p_{x,in}^{2}-p_{y}^{2}))(K(q_{in})p_{x,in}+l(q_{in})p_{y}))\frac{p_{y}%
}{Q_{y}(q_{in})}.
\end{align}
Now, by applying to equations (\ref{dx-dz py}) the estimate (\ref{est}) with
$n=1$ (one Picard iteration), we can restore from (\ref{dx-dz cut}) all the
formulas of lemma \ref{error_of_reflection} (we use (\ref{zamp0}) to restore
$p_{x}$ from $\tilde{p}$, and use (\ref{ener}) to determine $p_{y,out}$; note
also that, up to $O(M^{(r)})$-terms, the interval of integration is symmetric
by virtue of (\ref{nul}), so the integrals of odd functions of $p_{y}$ in the
right-hand-sides of (\ref{dx-dz cut}) are $O((M^{(r)})^{2})$).

\bibliographystyle{aaai-named}
\bibliography{ref03n}

\providecommand{\bysame}{\leavevmode\hbox to3em{\hrulefill}\thinspace}
\providecommand{\MR}{\relax\ifhmode\unskip\space\fi MR }
\providecommand{\MRhref}[2]{%
  \href{http://www.ams.org/mathscinet-getitem?mr=#1}{#2}
}
\providecommand{\href}[2]{#2}
\begin{thebibliography}{10}

\bibitem{Bal88}
P.~R. Baldwin, \emph{Soft billiard systems.}, Phys. D \textbf{29} (1988),
  no.~3, 321--342.

\bibitem{BaTo04}
P.~B{\'a}lint and I.~P. T{\'o}th, \emph{Mixing and its rate in `soft' and
  `hard' billiards motivated by the {L}orentz process}, Phys. D \textbf{187}
  (2004), no.~1-4, 128--135, Microscopic chaos and transport in many-particle
  systems.

\bibitem{Birk27}
G.~D. Birkhoff, \emph{Dynamical systems}, Amer. Math. Soc. Colloq. Publ.
  \textbf{9} (1927).

\bibitem{BDRR04}
S.~Bolotin, A.~Delshams, and R.~Ram{\'{\i}}rez-Ros, \emph{Persistence of
  homoclinic orbits for billiards and twist maps}, Nonlinearity \textbf{17}
  (2004), no.~4, 1153--1177.

\bibitem{Bu79}
L.~A. Bunimovich, \emph{On the ergodic properties of nowhere dispersing
  billiards}, Comm. Math. Phys. \textbf{65} (1979), no.~3, 295--312.

\bibitem{Bu01}
\bysame, \emph{Mushrooms and other billiards with divided phase space}, Chaos
  \textbf{11} (2001), no.~4, 802--808.

\bibitem{BuRe98}
L.~A. Bunimovich and J.~Rehacek, \emph{How high-dimensional stadia look like},
  Comm. Math. Phys. \textbf{197} (1998), no.~2, 277--301.

\bibitem{BuRe98f}
\bysame, \emph{On the ergodicity of many-dimensional focusing billiards}, Ann.
  Inst. H. Poincar\'e Phys. Th\'eor. \textbf{68} (1998), no.~4, 421--448,
  Classical and quantum chaos.

\bibitem{Chen04}
Y-C. Chen, \emph{Anti-integrability in scattering billiards}, Dyn. Syst.
  \textbf{19} (2004), no.~2, 145--159.

\bibitem{ChMar03}
N.~Chernov and R.~Markarian, \emph{Introduction to the ergodic theory of
  chaotic billiards}, second ed., Publica\c c\~oes Matem\'aticas do IMPA. [IMPA
  Mathematical Publications], Instituto de Matem\'atica Pura e Aplicada (IMPA),
  Rio de Janeiro, 2003, 24$\sp {\rm o}$ Col\'oquio Brasileiro de Matem\'atica.
  [24th Brazilian Mathematics Colloquium].

\bibitem{DFRR01}
A.~Delshams, Yu. Fedorov, and R.~Ram{\'{\i}}rez-Ros, \emph{Homoclinic billiard
  orbits inside symmetrically perturbed ellipsoids}, Nonlinearity \textbf{14}
  (2001), no.~5, 1141--1195.

\bibitem{DeRa96}
A.\ Delshams and R.~Ram\'irez-Ros, \emph{{P}oincar\'e-melnikov-arnold method
  for analytic planar maps}, Nonlinearity \textbf{9} (1996), 1--26.

\bibitem{Do96}
V.J. Donnay, \emph{Elliptic islands in generalized {S}inai billiards}, Ergod.
  Th. \& Dynam. Sys. \textbf{16} (1996), 975--1010.

\bibitem{do99}
V.J. Donnay, \emph{Non-ergodicity of two particles interacting via a smooth
  potential}, J Stat Phys \textbf{96} (1999), no.~5-6, 1021--1048.

\bibitem{DoLi91}
V.J. Donnay and C.~Liverani, \emph{Potentials on the two-torus for which the
  {H}amiltonian flow is ergodic.}, Commun. Math. Phys. \textbf{135} (1991),
  267--302.

\bibitem{DrR04}
Vladimir Dragovi{\'c} and Milena Radnovi{\'c}, \emph{Cayley-type conditions for
  billiards within {$k$} quadrics in {$\Bbb R\sp d$}}, J. Phys. A \textbf{37}
  (2004), no.~4, 1269--1276.

\bibitem{Gut90}
M.C. Gutzwiller, \emph{Chaos in classical and quantum mechanic},
  Springer-Verlag, New York, NY, 1990.

\bibitem{kfad01}
A.~Kaplan, N.~Friedman, M.~Andersen, and N.~Davidson, \emph{Observation of
  islands of stability in soft wall atom-optics billiards}, PHYSICAL REVIEW LET
  \textbf{87} (2001), no.~27, 274101--1--4.

\bibitem{Kn89}
A.~Knauf, \emph{On soft billiard systems.}, Phys. D \textbf{36} (1989), no.~3,
  259--262.

\bibitem{KzTr91}
V.~V. Kozlov and D.~V. Treshch{\"e}v, \emph{Billiards: A genetic introduction
  to the dynamics of systems with impacts}, American Mathematical Society,
  Providence, RI, 1991, Translated from the Russian by J. R. Schulenberger.

\bibitem{Kry79}
N.~S. Krylov, \emph{Works on the foundations of statistical physics}, Princeton
  University Press, Princeton, N.J., 1979, Translated from the Russian by A. B.
  Migdal, Ya. G. Sinai and Yu. L. Zeeman.

\bibitem{Ku76}
I.~Kubo, \emph{Perturbed billiard systems i the ergodicity of the motion of a
  particle in a compound central field}, Nagoya Math. J. \textbf{61} (1976),
  1--57.

\bibitem{KuMu81}
I.~Kubo and H.~Murata, \emph{Perturbed billiard systems {II} {B}ernoulli
  properties}, Nagoya Math. J. \textbf{81} (1981), 1--25.

\bibitem{Mar68}
J.~E. Marsden, \emph{Generalized {H}amiltonian mechanics: {A} mathematical
  exposition of non-smooth dynamical systems and classical {H}amiltonian
  mechanics}, Arch. Rational Mech. Anal. \textbf{28} (1967/1968), 323--361.
  \MR{MR0224935 (37 \#534)}

\bibitem{MarW01}
J.~E. Marsden and M.~West, \emph{Discrete mechanics and variational
  integrators}, Acta Numer. \textbf{10} (2001), 357--514.

\bibitem{PrSm00}
H.~Primack and U.~Smilansky, \emph{The quantum three-dimensional {S}inai
  billiard---a semiclassical analysis}, Phys. Rep. \textbf{327} (2000),
  no.~1-2, 107.

\bibitem{RKTu99}
V.~Rom{-}Kedar and D.~Turaev, \emph{Big islands in dispersing billiard-like
  potentials}, Physica D \textbf{130} (1999), 187--210.

\bibitem{SiNa04}
N.~Sim{\'a}nyi, \emph{Proof of the ergodic hypothesis for typical hard ball
  systems}, Ann. Henri Poincar\'e \textbf{5} (2004), no.~2, 203--233.

\bibitem{Sina63}
Ya.G. Sinai, \emph{On the foundations of the ergodic hypothesis for dynamical
  system of statistical mechanics}, Dokl. Akad. Nauk. SSSR \textbf{153} (1963),
  1261--1264.

\bibitem{Sina70}
\bysame, \emph{Dynamical systems with elastic reflections: Ergodic properties
  of scattering billiards}, Russian Math.\ Sur. \textbf{25} (1970), no.~1,
  137--189.

\bibitem{SiCh87}
Ya.G.\ Sinai and N.I. Chernov, \emph{Ergodic properties of some systems of
  two-dimensional disks and three-dimensional balls}, Uspekhi Mat.\ Nauk
  \textbf{42} (1987), no.~3(255), 153--174, 256, In Russian.

\bibitem{Smil95}
U.~Smilansky, \emph{Semiclassical quantization of chaotic billiards - a
  scattering approach}, Proceedings of the 1994 Les-Houches summer school on
  "Mesoscopic quantum Physics" (A.~Akkermans, G.~Montambaux, and J.L. Pichard,
  eds.), 1995.

\bibitem{Sz96}
D.~Sz\'asz, \emph{Boltzmann's ergodic hypothesis, a conjecture for centuries?},
  Studia Sci. Math. Hungar. \textbf{31} (1996), no.~1--3, 299--322.

\bibitem{Tab95}
S.~Tabachnikov, \emph{Billiards}, Panor. Synth. (1995), no.~1, vi+142.

\bibitem{TuRK98}
D.~Turaev and V.~Rom{-}Kedar, \emph{Islands appearing in near-ergodic flows},
  Nonlinearity \textbf{11} (1998), no.~3, 575--600.

\bibitem{turk03}
D.~Turaev and V.~Rom-Kedar, \emph{Soft billiards with corners}, J. Stat. Phys.
  \textbf{112} (2003), no.~3--4, 765--813.

\bibitem{Ves91}
A.~P. Veselov, \emph{Integrable mappings}, Uspekhi Mat. Nauk \textbf{46}
  (1991), no.~5(281), 3--45, 190.

\bibitem{Wo86}
M.~Wojtkowski, \emph{Principles for the design of billiards with nonvanishing
  lyapunov exponents}, Comm. Math. Phys. \textbf{105} (1986), no.~3, 391--414.

\bibitem{Wo90}
\bysame, \emph{Linearly stable orbits in {$3$}-dimensional billiards}, Comm.
  Math. Phys. \textbf{129} (1990), no.~2, 319--327.

\end{thebibliography}

\end{document}